\documentclass[10pt,journal,compsoc]{IEEEtran}

\pdfoutput=1
\usepackage{amsfonts}
\usepackage{algorithm}
\usepackage{algorithmic}
\usepackage{amsmath}
\usepackage{amssymb}
\usepackage{amsfonts}
\usepackage{bm}
\usepackage{color}
\usepackage{comment}
\usepackage{graphicx}
\usepackage{multirow}
\usepackage{subfigure}
\usepackage{url}

\newtheorem{definition}{Definition}

\ifCLASSOPTIONcompsoc

\else

\fi

\ifCLASSINFOpdf

\else

\fi

\begin{document}

\title{ A Hierarchical Attention Model for Social Contextual Image Recommendation}

\author{Le Wu~\IEEEmembership{Member,~IEEE}, Lei Chen, Richang Hong,~\IEEEmembership{Member,~IEEE},
Yanjie Fu, \\
Xing Xie,~\IEEEmembership{Senior Member,~IEEE}, Meng Wang,~\IEEEmembership{Senior Member,~IEEE}

\IEEEcompsocitemizethanks{

\IEEEcompsocthanksitem  L.~Wu, L.~Chen, R.~Hong, M.~Wang are with the School of Computer and Information, Hefei University of Technology,
Hefei, Anhui 230009, China.
\protect \\Emails: \{lewu.ustc, chenlei182979,hongrc.hfut,eric.mengwang\}@gmail.com.

\IEEEcompsocthanksitem Y.~Fu is with the Department of Computer Science,
University of Missouri-Rolla, Rolla, MO, USA. \protect Email: fuyan@mst.edu.

\IEEEcompsocthanksitem X.~Xie is with Microsoft Research, Beijing, China.
\protect \newline Email: xingx@microsoft.com.
}

}

\markboth{IEEE TRANSACTIONS ON KNOWLEDGE AND DATA ENGINEERING}
{Shell \MakeLowercase{\textit{et al.}}: Bare Advanced Demo of IEEEtran.cls for Journals}

\IEEEtitleabstractindextext{
\begin{abstract}



Image based social networks are among the most popular social networking services in recent years. With tremendous images uploaded everyday, understanding users' preferences on user-generated images and making recommendations have become an urgent need.  In fact, many hybrid models have been proposed to fuse various kinds of side information~(e.g., image visual representation, social network) and user-item historical behavior for enhancing recommendation performance. However, due to the unique characteristics of the user generated images in social image platforms, the previous studies failed to capture the complex aspects that influence users' preferences in a unified framework. Moreover, most of these hybrid models relied on predefined weights in combining different kinds of information, which usually resulted in sub-optimal recommendation performance.  To this end, in this paper, we develop a hierarchical attention model for social contextual image recommendation. In addition to basic latent user interest modeling in the popular matrix factorization based recommendation, we identify three key aspects~(i.e., upload history, social influence,  and owner admiration) that affect each user's latent preferences, where each aspect summarizes a contextual factor from the complex relationships between users and images.
After that, we design a hierarchical attention network that  naturally mirrors the hierarchical relationship~(elements in each aspects level, and the aspect level) of users' latent interests with the identified key aspects. Specifically, by taking embeddings from state-of-the-art deep learning models that are tailored for each kind of data, the hierarchical attention network could learn to attend differently to more or less content. Finally, extensive experimental results on real-world datasets clearly show the superiority of our proposed model.

\end{abstract}


}


\maketitle


\IEEEpeerreviewmaketitle

\ifCLASSOPTIONcompsoc
\IEEEraisesectionheading{\section{Introduction}\label{sec:introduction}}
\else

\vspace{-0.2cm}
\section{Introduction} \label{sec:intro}
\fi

There is an old saying ``a picture is worth a thousand words". When it comes to social media, it turns out that visual images are growing much more popularity to attract users~\cite{MM2018beyond}. Especially with the increasing adoption of smartphones, users could easily take qualified images and upload them to various social image platforms to share these visually appealing pictures with others. Many image-based social sharing services have emerged, such as \emph{Instagram}\footnote{https://www.instagram.com},
\emph{Pinterest}\footnote{https://www.pinterest.com},
and \emph{Flickr}\footnote{https://www.flickr.com}.
With hundreds of millions of images uploaded everyday, image recommendation has become an urgent need to deal with the image overload problem. By providing personalized image suggestions to each active user in image recommender system, users gain more satisfaction for platform prosperity. E.g., as reported by \emph{Pinterest}, image recommendation powers over 40\% of user engagement of this social platform~\cite{WWW2017pinterest}.


Naturally, the standard recommendation algorithms  provide a direct solution for the image recommendation task~\cite{TKDE2005toward}. For example,  many classical latent factor based Collaborative  Filtering~(CF) algorithms in recommender systems could be applied to deal with user-image interaction  matrix~\cite{KDD2008factorization,UAI2009bpr,KDD2008factorization}.
Successful as they are, the extreme data sparsity of the user-image interaction behavior limits the recommendation performance~\cite{TKDE2005toward,KDD2008factorization}. On one hand, some recent works proposed to enhance recommendation performance with visual contents learned from a (pre-trained) deep neural network~\cite{AAAI2016vbpr,WWW2017Visual,SIGIR2017attentive}. On the other hand, as users perform image preferences in social platforms, some social based recommendation algorithms utilized the social influence among users to alleviate data sparsity for better recommendation~\cite{WSDM2011recommender,TKDE2014scalable,KDD2008influence}.  In summary, these studies partially solved the data sparsity issue of social-based image recommendation. Nevertheless, the problem of how to better exploit the unique characteristics of the social image platforms in a holistical way to enhance recommendation performance is still under explored.


\begin{small}
\begin{figure*} [htb]
 \begin{center}
 \vspace{-0.3cm}
\includegraphics[width=160mm]{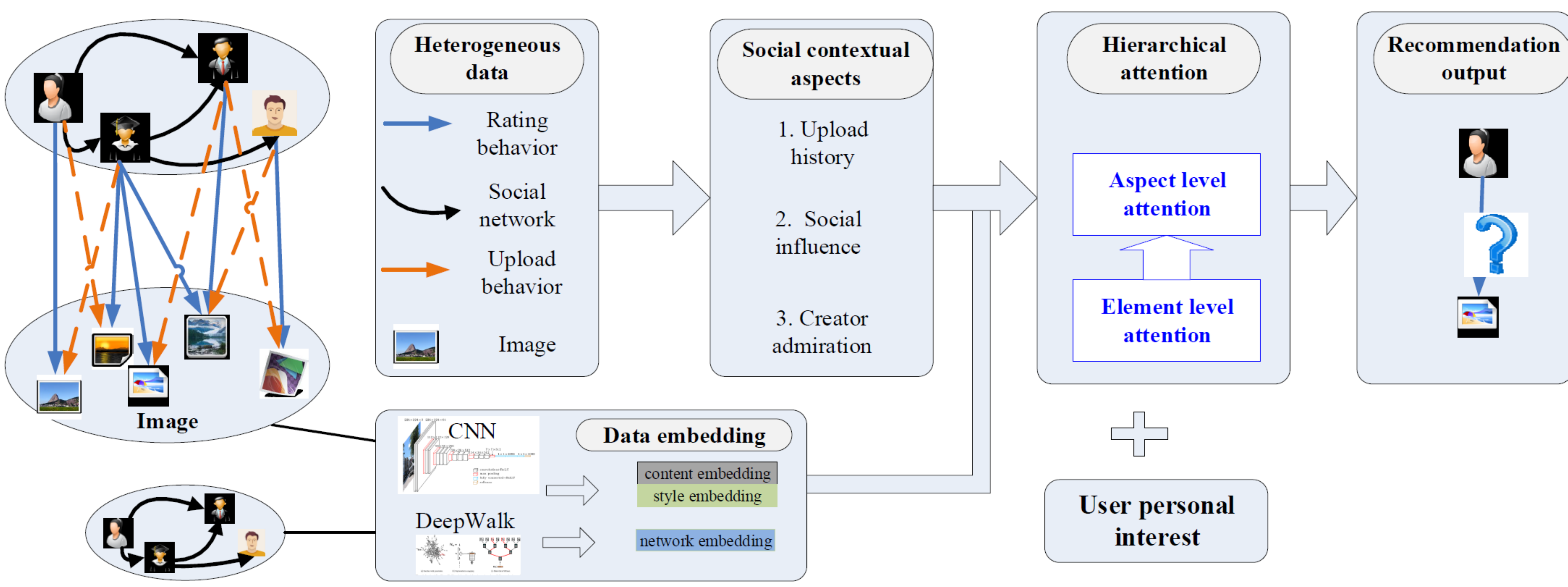}
  \end{center}
  \vspace{-0.2cm}
  \caption{\small{An overall framework of social contextual image recommendation, where the left part shows the data characteristics of the platform, and the right part shows our proposed model.} } \label{fig:data}
  \vspace{-0.4cm}
\end{figure*}
\end{small}

In this paper, we study the problem of understanding users' preferences for images and recommending images in social image based platforms. Fig.~\ref{fig:data} shows an example of a typical social image application. Each image is associated with  visual information. Besides showing likeness to images, users are also creators of these images with the upload behavior. In addition, users connect with others to form a social network to share their image preferences. The rich heterogeneous contextual data provides valuable clues to infer users' preferences to images.
Given rich heterogeneous contextual data,  the problem of how to summarize  the heterogeneous \emph{social contextual aspects} that influence users' preferences to these highly subjective content is still unclear.
What's more, in the preference decision process, different users care about different social contextual aspects for their personalized image preference. E.g. \emph{Lily} likes images that are similar to her uploaded images,  while \emph{Bob} is easily swayed by social neighbors to present similar preference as her social friends. In other words, the unique user preference for balancing these complex social contextual aspect makes the recommendation problem more challenging.

To address the challenges mentioned above, in this paper, we design a hierarchical attention model for social image recommendation. The proposed model is built on the popular latent factor based models, which assumes users and items could be projected in a low latent space~\cite{NIPS2008probabilistic}. In our proposed model, for each user, in addition to basic latent user interest vector, we identify three key aspects~(i.e., upload history, social influence and owner admiration) that affect each user's preference, where each aspect summarizes a contextual factor from the complex relationships between users and images. Specifically, the upload history aspect summarizes each user's uploaded images to characterize her interest. The social influence aspect characterizes the influence from the social network structure, and the owner admiration aspect depicts the influence from the uploader of the recommended image. The three key aspects are combined to form the auxiliary user latent embedding. Furthermore, since not all aspects are equally important for personalized image recommendation, we design a hierarchical attention structure that attentively weight different aspects for each user's auxiliary embedding. The proposed hierarchical structure aims at capturing the following two distinctive characteristics. First, as social contextual recommendation naturally exhibits the hierarchical structure~(various elements from each aspect, and the three aspects of each user), we likewise construct user interest representation with a hierarchical structure. In the hierarchical structure, we first build auxiliary aspect  representations of each user, and then aggregate the three aspect representations into an auxiliary user interest vector. Second, as different elements within each aspect, and different aspects are differentially informative  for each user in the recommendation process, the hierarchical attention network builds two levels of attention mechanisms that apply at the element level and the aspect level.


We summarize the contributions of this paper as follows:

\begin{enumerate}
  \item We study the problem of image recommendation in social image based platforms. By considering the uniqueness of these platforms, we identify three social contextual aspects that affect users' preferences from heterogeneous data sources.
  \item We design a hierarchical attention network to model the hierarchical structure of social contextual recommendation. In the attention networks, we feed  embeddings from state-of-the-art deep learning models that are tailored for each kind of data into the attention networks. Thus, the attention networks could learn to attend differently based on the rich contextual information for user interest modeling.
  \item We conduct extensive experiments on real-world datasets. The experimental results clearly show the effectiveness of our proposed model.
\end{enumerate}




\section{Related Work}


We summarize the related work in the following four categories.

\textbf{General Recommendation.} Recommender systems could be classified into three categories: content based methods, \emph{C}ollaborative \emph{F}iltering~(CF) and the hybrid models~\cite{TKDE2005toward}. Among all models for building recommender systems, latent factor based models from the CF category are among the most popular techniques due to their relatively high performance in practice~\cite{UAI2009bpr,NIPS2008probabilistic,rendle2012factorization}. These latent factor based models decomposed both users and items in a low latent space, and the preference of a user to an item could be approximated as the inner product between the corresponding user and item latent vectors. In the real-world applications, instead of the explicit ratings, users usually implicitly express their opinions through action or inaction. Bayesian Personalized Ranking~(BPR) is such a popular latent factor based model that deals with the implicit feedback~\cite{UAI2009bpr}. Specifically, BPR optimized a pairwise based ranking loss, such that the observed implicit feedbacks are preferred to rank higher than that of the unobserved ones. As users may simultaneously express their opinions with several kinds of feedbacks~(e.g., click behavior, consumption behavior). SVD++ is proposed to incorporate users' different feedbacks by extending the classical latent factor based models, assuming each user's latent factor is composed of a base latent factor, and an auxiliary latent factor that can be derived from other kinds of feedbacks~\cite{KDD2008factorization}. Due to the performance improvement and extensibility of SVD++, it is widely studied to incorporate different kinds of information, e.g., item text~\cite{SIGIR2017autosvd++}, multi-class preference of users~\cite{pan2017collaborative}.

\textbf{Image Recommendation.}
In many image based social networks, images are associated with rich context information, e.g., the text in the image, the hashtags. Researchers proposed to apply factorization machines for image recommendation by considering the rich context information~\cite{MM2016context}. Recently, deep Convolutional Neural Networks(CNNs) have been successfully applied to analyzing visual imagery by automatic image representation in the modeling process~\cite{NIPS2012imagenet}. Thus, it is a natural idea to leverage visual features of CNNs to enhance image recommendation performance~\cite{AAAI2016vbpr,CVPR2016comparative, Recsys2016Vista,SIGIR2017attentive}. E.g., VBPR is an extension of BPR for image recommendation, on top of which it learned an additional visual dimension from CNN that modeled users' visual preferences~\cite{AAAI2016vbpr}. There are some other image recommendation models that tackled the temporal dynamics of users' preferences to images over time~\cite{Recsys2016Vista}, or users' location preferences for image recommendation~\cite{WSDM2018neural,WWW2017Visual,WSDM2018neural}. 
As well studied in the computer vision community,  in parallel to the visual content information from deep CNNs, images convey rich style information. Researchers showed that many brands post images that show the philosophy and lifestyle of a brand~\cite{MM2018beyond}, images posted by users also reflect users' personality~\cite{MM2017personality}. Recently, Gatys et al. proposed a new model of extracting image styles based on the feature maps of convolutional neural networks~\cite{NIPS2015texture}. The proposed model showed high perceptual quality for extracting image style,  and has been successfully  applied to related tasks, such as image style transfer~\cite{gatys2016image}, and high-resolution image stylisation~\cite{CVPR2017controlling}.  We argue that the visual image style also plays a vital role for evaluating users' visual experience in recommender systems. Thus, we leverage both the image content and the image style for recommendation.

\textbf{Social Contextual Recommendation.}
Social scientists have long converged that a user's preference is similar to or influenced by  her social connections, with the social theories of  homophily and social influence~\cite{KDD2008influence}. With the prevalence of social networks, a popular research direction is to leverage the social data to improve recommendation performance~\cite{WSDM2011recommender,Recsys2010matrix,TKDE2014scalable,TSMC2019Social}. E.g., Ma et al. proposed a latent factor based model with social regularization terms for recommendation~\cite{WSDM2011recommender}. Since most of these social recommendation tasks are formulated as non-convex optimizing problems, researchers have designed an unsupervised deep learning model to initialize model parameters for better performance~\cite{TNNLS2017deep}. Besides, ContextMF is proposed to fuse the individual preference and interpersonal influence with auxiliary text content information from social networks~\cite{TKDE2014scalable}. As the implicit influence of trusts and ratings are valuable for recommendation, TrustSVD is proposed to incorporate the influence of trusted users on the prediction of items for an active user~\cite{TKDE2016novel}. The proposed technique extended the SVD++ with social trust information. Social recommendation has also been considered with social circle~\cite{TKDE2014personalized}, online social recommendation~\cite{TKDE2016user}, social network evolution~\cite{TKDE2017modeling}, and so on.

Besides, as the social network could be seen as a graph, the recent surge of network embedding is also closely related to our work~\cite{TKDE2018survey}. Network embedding models encode the graph structural information into a low latent space, such that each node is represented as an embedding in this latent space. Many network embedding models have been proposed~\cite{KDD2014deepwalk,tang2015line,wang2016structural,velickovic2017graph}. The network embedding could be used for the attention networks. We distinguish from these works as the focus of this paper is not to advance the sophisticated network embedding models. We put emphasis on how to enhance recommendation performance by leveraging various data embeddings.


\textbf{Attention Mechanism.}  Neural science studies have shown that people focus on specific parts of the input rather than using all available information~\cite{PAMI1998model}. Attention mechanism is such an intuitive idea that automatically models and selects the most pertinent piece of information, which learns to assign attentive weights for a set of inputs, with higher~(lower) weights indicate that the corresponding inputs are more informative to generate the output. Attention mechanism is widely used in many neural network based tasks, such as machine translation~\cite{Neural2014neural} and image captioning~\cite{ICML2015show}. Recently, the attention mechanism is also widely used for recommender systems~\cite{he2018nais,xiao2017attentional,sun2018attentive,seo2017interpretable}.
Given the classical collaborative filtering scenario with user-item interaction behavior, NAIS extended the classical item based recommendation models by distinguishing the importance of different historical items in a user profile~\cite{he2018nais}. With users' temporal behavior, the attention networks were proposed to learn which historical behavior is more important for the user's current temporal decision~\cite{liu2018stamp,loyola2017modeling}.
A lot of attention based recommendation models have been developed to better exploit the auxiliary information to improve recommendation performance. E.g., ANSR is proposed with a social attention module to learn adaptive social influence strength for social recommendation~\cite{sun2018attentive}.  Given the review or the text of an item, attention networks were developed to learn informative sentences or words for recommendation~\cite{gong2016hashtag,seo2017interpretable}. While the above models perform the standard vanilla attention to learn to attend on a specific piece of information, the co-attention mechanism is concerned to learn attention weights from two sequences~\cite{hu2018leveraging,zhang2017hashtag,tay2018multi}. E.g., in the hashtag recommendation with both text and image information, the co-attention network is designed to learn which part of the text is distinctive for images, and simultaneously the important visual features for the text~\cite{zhang2017hashtag}.
Besides, researchers have made a comprehensive survey the attention based recommendation models~\cite{zhang2019deep}.
In  some real-world applications, there exists hierarchical structure among the data, several pioneering works have been proposed to deal with this kind of relationship~\cite{NAACL2016hierarchical,li2015hierarchical}. E.g., a hierarchical attention model is proposed to model the hierarchical relationships of word, sentence and document for document classification~\cite{NAACL2016hierarchical}. Our work borrows ideas from the attention mechanism, and we extend this idea by designing a hierarchical structure to model the complex social contextual aspects that influence users' preferences. Nevertheless, different from the natural hierarchical structure of words, sentences and documents in natural language processing, the hierarchial structure that influences a user's decision from complex heterogenous data sources is summarized by our proposed model. Specifically, our proposed model has a two-layered hierarchical structure with the bottom layer attention network that summarizes each aspect from the various elements of this aspect. By taking the output of each aspect from the bottom layer, the top-layer attention network learns the importance of the three aspects.

The work that is most similar to ours is the \emph{A}ttentive \emph{C}ollaborative \emph{F}iltering~(ACF) for image and video recommendation~\cite{SIGIR2017attentive}. By assuming there exists item level and component level implicitness that  underlines a user's preference, an attention based recommendation model is proposed with the component level attention and the item level attention. Our work borrows the idea of applying attention mechanism for recommendation, and it differs from ACF and previous works from both the research perspective and the application point. From the technical perspective, we model the complex social contextual aspects of users' interests from heterogeneous data sources in a unified recommendation model. In contrast, ACF only leverages the image~(video) content information. From the application view, our proposed model could benefit researchers and engineers in related areas when heterogeneous data are available.

\section{Heterogeneous Data Embedding and Problem Definition}
In a social image platform, there are a set of users {\small $U$~($|U|\!=\!M$)} and a set of images {\small$V$~($|V|\!=\!N$)}.  Besides \emph{rating images} as standard recommender systems, users also perform two kinds of behaviors: \emph{uploading images} and \emph{building social links}. We represent users' three kinds of behaviors with three matrices: a rating matrix {\small $\mathbf{R}\in \mathbb{R}^{M\times N}$}, an upload matrix {\small $\mathbf{L}\in \mathbb{R}^{N\times M}$}, and a social link matrix {\small $\mathbf{S}\in \mathbb{R}^{M\times M}$}. Each element $r_{ai}$ in the rating matrix {\small $\mathbf{R}$} represents the implicit rating preference of user $a$ to image $i$, with {\small$r_{ai}\!=\!1$} denotes user $a$ likes image $i$, otherwise it equals 0. {\small$s_{ba}\!=\!1$} if user $a$ follows~(connects to) user $b$, otherwise it equals 0. If the social platform is undirected, $a$ connects to $b$ means $s_{ab}\!=\!1$ and $s_{ba}\!=\!1$. We use $\mathbf{s}_a=[s_{1a},s_{2a},...,s_{Ma}]$  to denote the social connections of $a$, i.e., the $a$-th column of {\small$\mathbf{S}$}. Please note that different from traditional social networking platforms~(e.g., the social movie sharing platform), users in these platforms are both image consumers~(i.e., reflected in the rating behavior) and image creators~(reflected in the upload behavior). Each element $l_{ia}$ in the upload matrix {\small$\mathbf{L}$} denotes whether the image $i$ is uploaded~(created) by user $a$. In other words, if $a$ is the creator of image $i$, then {\small$l_{ia}\!=\!1$}, otherwise it equals 0. Since each image can be uploaded by only one user, we have {\small$\sum_{a=1}^M l_{ia}\!=\!1$}. For ease of explanation, we use $C_i$ to denote the creator of image $i$. And the image upload history of $a$ is denoted as $\mathbf{l}_a$, i.e., the $a$-th column of {\small$\mathbf{L}$}. Without confusion,  we use $a,b,c$ to represent users and $i,j,k$ to denote items.

\begin{small}
\begin{figure} [htb]
 \begin{center}
\includegraphics[width=88mm]{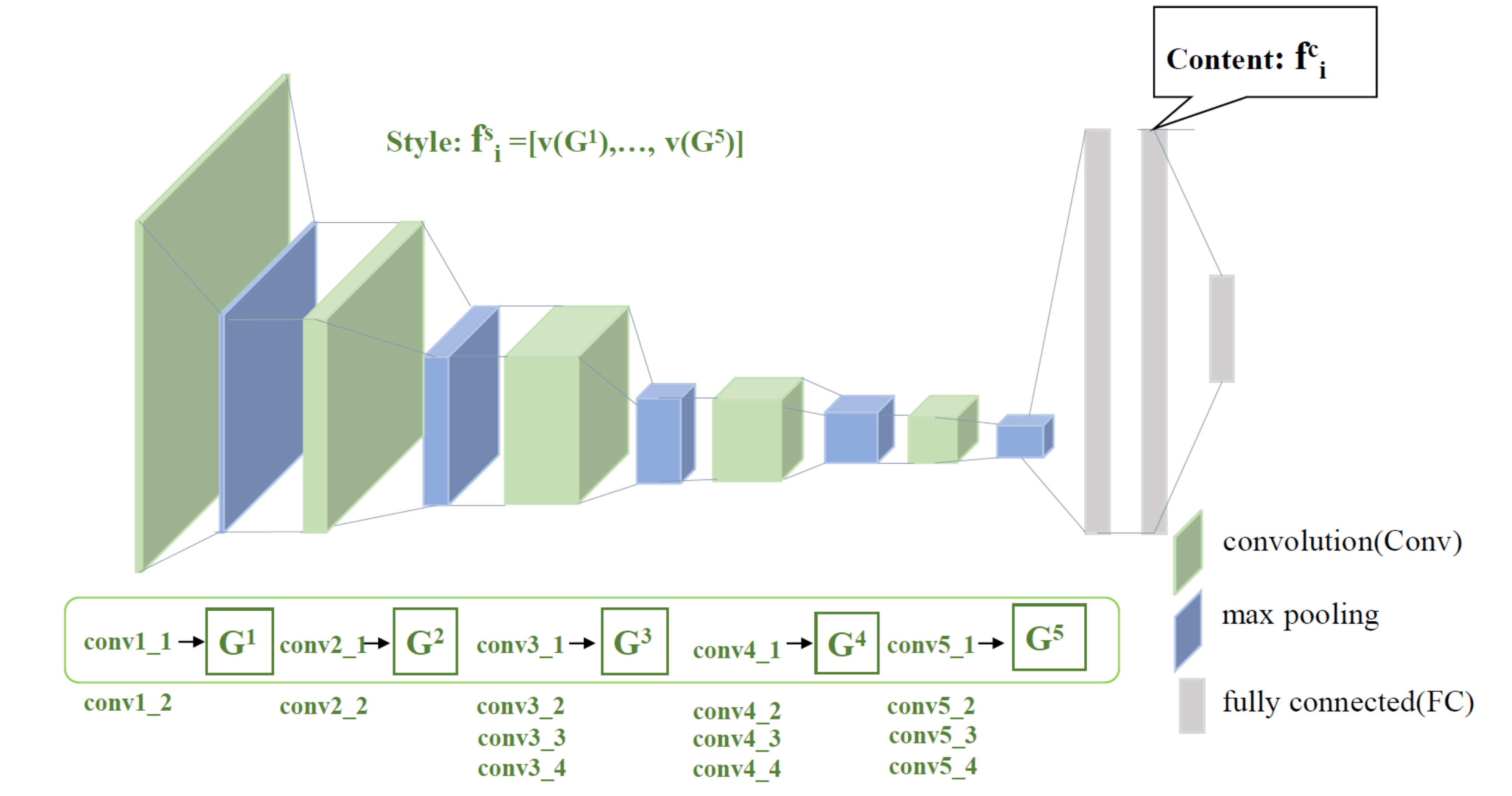}
  \end{center}
  \vspace{-0.4cm}
  \caption{\small{The image embedding process. For each image $i$, the original image is passed through a VGG19 network. We use the vector of the last connected layer, i.e., $\mathbf{f}^c_i$ as its content representation. The Gram matrices $\mathbf{G}^l$ on the feature responses of a number of layers are computed. We concatenate the vectorized representations of the typical Gram matrix sequences as the image style representation, i.e., $\mathbf{f}^s_i$.} } \label{fig:img_embed}
  \vspace{-0.2cm}
\end{figure}
\end{small}

\subsection{Heterogeneous Data Embedding}
Since there are heterogeneous data sources in this platform, it is natural to adopt the state-of-the-art data embedding techniques to preprocess the social network  {\small $\mathbf{S}$} and the visual images.  The learned embeddings are easier to be exploited by the following proposed model than directly dealing with the heterogeneous data sources.  Next, we would first briefly introduce the  embedding models for the social network and the visual images, and then give the problem definition. Please note that, the problem of how to design sophisticated network embedding techniques, and the visual image features are well researched. Since the focus of this paper is not to advance these topics, we adopt state-of-the-art models and put emphasis on enhancing the recommendation performance with the rich social contextual information.

For the social network  {\small $\mathbf{S}$},  the social embedding part tries to learn the distributed representation of each user in the  social network {\small$\mathbf{S}$},  which encodes social relations in a continuous vector space. Since the focus of this paper is not to design more sophisticated models for network embedding, we exploit Deepwalk~\cite{KDD2014deepwalk} for social embedding as it is time-efficient and shows high performance in many network based applications. Deepwalk takes {\small$\mathbf{S}$} as input and outputs the social latent representation {\small$\mathbf{E}\in\mathbb{R}^{d\times M}$}, with the a-th column  $\mathbf{e}_a$ denotes the latent representation of user $a$.

For each image, it provides rich information including its content as well as its style.
Traditionally, convolution neural networks have enjoyed great success for learning useful image visual content features in recent years~\cite{NIPS2012imagenet,vgg2014very}. We choose \emph{VGG19} for visual content feature extraction as it is a state-of-the-art convolutional neural network architecture that shows powerful capability to capture the image semantics~\cite{vgg2014very}. As commonly adopted by many works, we use the 4096 dimensional representation in the last connected layer in VGG19 as the visual content representation, i.e., each image $i$'s visual feature $\mathbf{f}^c_i$ has 4096 dimensions~\cite{AAAI2016vbpr,KDD2016collaborative}.

Besides image content representation, the image style also plays a vital role for users' visual experience. When users browse images in social platforms, their preferences are not only decided by ``what is the content of the image?", but also ``does the style of the image meets my preference?''. To this end, for each image $i$, besides its content representation $\mathbf{f}^c_i$, we propose to borrow state-of-the-art image style representation models to capture its style representation $\mathbf{f}^s_i$. We choose a popular image style representation method proposed by Gatys et al.~\cite{NIPS2015texture}. This method has shown high perceptual quality and is widely used in many image-style based tasks~\cite{gatys2016image,CVPR2017controlling}. This style describing model is based on the powerful feature spaces learned by convolutional neural networks, with the assumption that the styles are agnostic to the spatial information in the image hidden representations. With the trained VGG19 architecture, suppose a layer $l$ has $N_l$ distinct filter feature maps, each of which is vectorized into a size of $M_l$. Let {\small$\mathbf{B^l}\in \mathbb{R}^{N_l\times M_l}$} denotes the filter at layer $l$, with $b^l_{jk}$ is the activation of the j-th filter at position $k$. A summary \emph{Gram} statistic is proposed to discard the spatial information in the feature maps by computing their relations as:

\begin{small}
\begin{equation}
g^l_{ij}=\sum_{k} b^l_{ik}b^l_{jk}
\end{equation}
\end{small}

\noindent where {\small$\mathbf{G^l}\in \mathbb{R}^{N_l\times N_l}$} is the Gram matrix, with $g^l_{ij}$ denotes the correlation between feature map $i$ and $j$ in layer $l$.  Naturally, the set of Gram matrices {\small${\mathbf{G^1}, \mathbf{G^2},...,\mathbf{G^L}}$} from different layers of VGG19 provides descriptions of the image style. In practice, researchers found that the style representations on layers `con1\_1', `conv2\_1', `con3\_1',  `con4\_1' and  `con5\_1' can well represent the textures of an image~\cite{gatys2016image,CVPR2017controlling}. As the sizes of these Gram matrices are very large, we downsample each Gram matrix into a fixed size of $32\times 32$, and then concatenate the vector representation of the downsampled Gram matrices of the five layers. Since there are 5 Gram matrices and each
each vectorized Gram matrix has 1024 dimensions, the style representation $\mathbf{f}^s_i$ of image $i$  has 5120 dimensions.

\subsection{Problem Definition}
Given the social matrix {\small$\mathbf{S}$} and upload matrix {\small$\mathbf{L}$}, we identify three key social contextual aspects, i.e.,  social influence, upload history, and the creator admiration that may influence users' preferences. Specifically, the \emph{social influence} aspect from each user $a$'s social network structure $\mathbf{s}_a$ is well recognized as an important factor in the recommendation process~\cite{WSDM2011recommender,TKDE2014scalable}. The social influence states that, each active user is influenced by her social connections, leading to the similar preferences between social connections~\cite{KDD2008influence}. Besides, for each user-item pair $(a,i)$, we could get an upload history list $\mathbf{l}_a$ of user $a$, and the creator $C_i$ of image $i$ from the upload matrix {\small$\mathbf{L}$}. Based on this observation, we design the two contextual aspects in users' preference decision process: an \emph{upload history} aspect that explains the consistency between her upload history $\mathbf{l}_{a}$ and her preference for images, and the \emph{creator admiration} aspect that shows the admiration from the creator $C_i$. These three contextual aspects characterise each user's implicit feedback to images from various contextual situations from the heterogeneous social image data.
Now, we define the social contextual image recommendation problem as:

\begin{definition}  \label{def:prob_def}[\textbf{PROBLEM DEFINITION}] Given the user rating matrix {\small$\mathbf{R}$}, the upload matrix {\small$\mathbf{L}$}, and the social network {\small$\mathbf{S}$} in a social image platform, with the social embedding $\mathbf{e}_a$ of each user $a$, and the content representation $\mathbf{f}^c_i$ and style representation $\mathbf{f}^s_i$ of each image $i$,  the social contextual recommendation task aims at: predicting each user $a$'s unknown preference for image $i$ with the three social contextual aspects~($\mathbf{s}_a$, $\mathbf{l}_a$, {\small$C_i$}) and the heterogeneous data embeddings~($\mathbf{e}_a$, $\mathbf{f}^c_i$ and $\mathbf{f}^s_i$) as $g(a, i, \mathbf{s}_a, \mathbf{l}_a, {\small C_i}, \mathbf{e}_a, \mathbf{f}^c_a, \mathbf{f}^s_a)$;
\end{definition}

\noindent Specifically, in the above definition, $\mathbf{s}_a$, $\mathbf{l}_a$, and {\small$C_i$} denotes the inputs of the three social contextual aspects, i.e., upload history aspect, social influence aspect and the creator admiration aspect.

In the following of this paper, we use bold capital letters to denote matrices, and small bold letters to denote vectors. For any matrix~(e.g., social graph {\small$\mathbf{S}$}), its $i$-th column vector is denoted as the corresponding small letter with a subscript index $i$~(e.g., the $i$-th column of {\small$\mathbf{S}$} is denoted as $\mathbf{s}_a$). We list some mathematical notations in Table~\ref{tab:math_notations}.

\begin{table}[htb] \centering
\caption{\small{Mathematical Notations}}\label{tab:math_notations}
\vspace{-0.2cm}
\begin{footnotesize}
\begin{tabular}{|l|l|}  \hline
Notations & Description \\ \hline \hline
U &  userset, $|U|=M$ \\ \hline
V &  imageset, $|V|=N$   \\ \hline
a,b,c,u & user \\ \hline
i,j,k,v & image\\ \hline
$\mathbf{R}\in \mathbb{R}^{M\times N}$  & rating matrix, with $r_{ai}$ denotes\\ & whether $a$ likes image $i$  \\ \hline
$\mathbf{S}\in \mathbb{R}^{M\times M}$  & social network matrix, with $s_{ba}$ denotes\\ & whether $a$ follows $b$\\ \hline
$\mathbf{L}\in \mathbb{R}^{N\times M}$  & upload matrix, with $l_{ia}$ denotes\\ & whether $a$ uploads image $i$ \\ \hline
$\mathbf{s}_a\in \mathbb{R}^{M}$ & the $a$-th column of $\mathbf{S}$,\\&which denotes the social connections of $a$ \\ \hline
$\mathbf{l}_a\in \mathbb{R}^{N}$ & the $a$-th column of $\mathbf{L}$, \\&which denotes the uploaded history of $a$\\ \hline
$C_i\in U$ & the creator~(owner) of image $i$, $C_i=[a: L_{ai}=1]$ \\ \hline
$\mathbf{e}_a$ & the social embedding of user $a$ from \\& social embedding matrix {\small$\mathbf{E}\in \mathbb{R}^{d\times M}$}\\ \hline
$\mathbf{f}^c_i$ &  the visual content representation of image $i$   \\ \hline
$\mathbf{f}^s_i$ &  the visual style representation of image $i$  \\ \hline
\end{tabular}
\end{footnotesize}
\vspace{-0.2cm}
\end{table}

\begin{small}
\vspace{-0.3cm}
\begin{figure*} [htb]
 \begin{center}
\includegraphics[width=140mm]{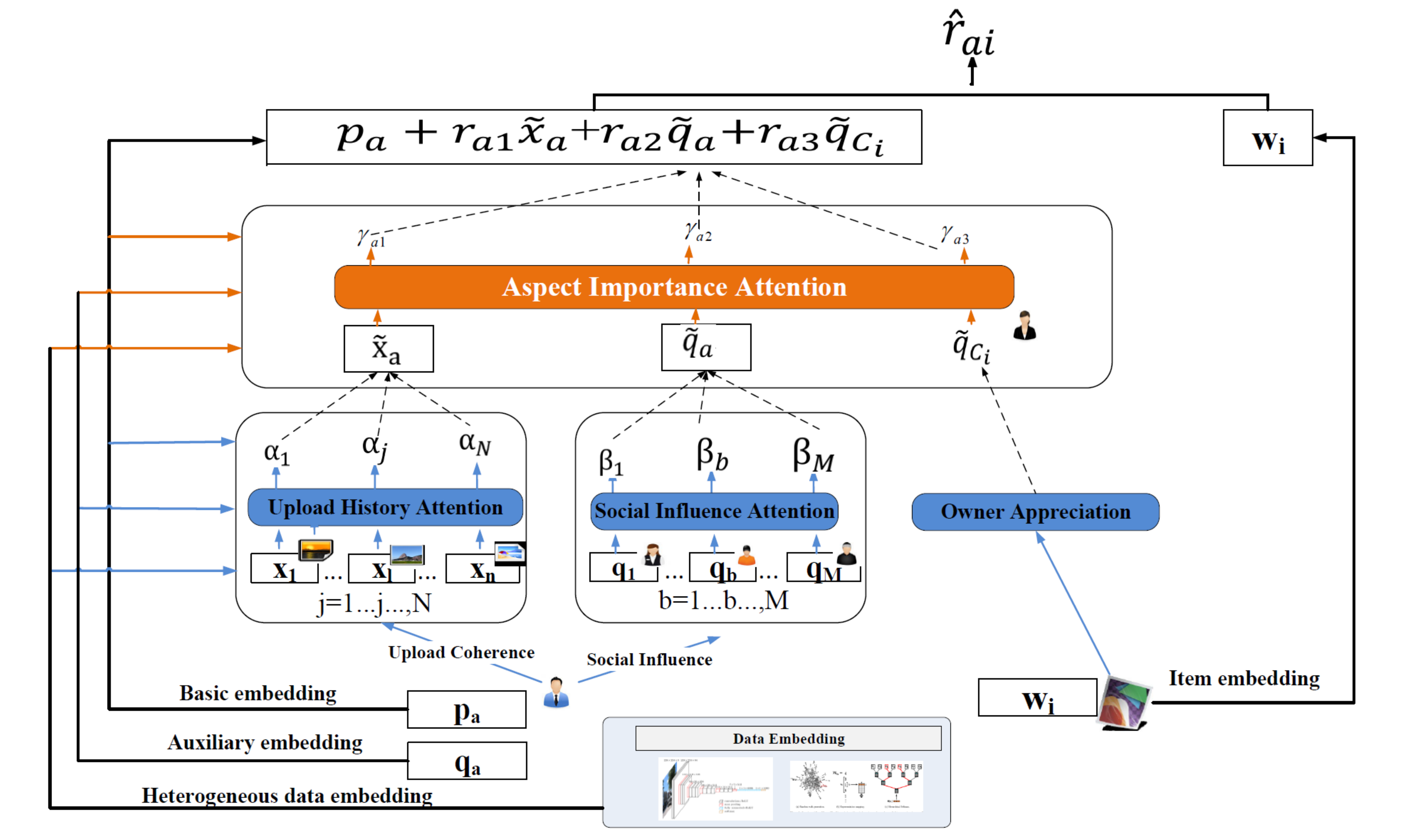}
  \end{center}
  \caption{\small{The overall architecture of the proposed HASC model.} } \label{fig:gra_framework}
  \vspace{-0.5cm}
\end{figure*}
 \vspace{-0.2cm}
\end{small}

\section{The Proposed Model}
In this section, we present our proposed \emph{H}ierarchical \emph{A}ttentive \emph{S}ocial \emph{C}ontextual recommendation~(HASC) model for image recommendation.

As shown in Fig.~\ref{fig:gra_framework}, HASC is a hierarchical neural network that models users' preferences for to unknown images from two attention levels with social contextual modeling. The top layered attention network depicts the importance of the three contextual aspects~(i.e., upload history, social influence and creator admiration) for users' decision, which is derived from the bottom layered attention networks that aggregate the complex elements within each aspect. Given a user $a$ and  an image $i$ with three identified social contextual aspects, we use $\gamma_{al}$~($l=1, 2, 3$) to denote $a$'s attentive degree for aspect $l$ on the top layer~(denoted as the aspect importance attention with orange part in the figure). A large attentive degree denotes the current user cares more about this aspect in image recommendation process. Besides, as there are various elements within the upload history context $\mathbf{l}_a$ and social influence context $\mathbf{s}_a$. We use $\alpha_{aj}$ to denote $a$'s preference degree for image $j$ in the upload history context {\small$l_a$}~($l_{ja}=1$), with a larger value of $\alpha_{aj}$ indicates that $a$'s current interest is more coherent with uploaded image $j$ by user $a$. Similarly, we use $\beta_{ab}$ to denote  the influence strength of the $b$ to $a$ in social neighbor context {$\mathbf{s}_a$}~($s_{ba}\!=\!1$), with a larger value of $\beta_{ab}$ indicates that $a$ is more likely to be influenced by $b$.
Please note that, for each user $a$ and image $i$, different from the upload history aspect and the social influence aspect, the creator admiration aspect is composed of one element $C_i$ ~(the creator). Thus, this aspect does not have any sub layers and it is directly sent to the top layer. We use three attention sub-networks to learn these attentive scores in a unified model. 

\textbf{Objective Prediction Function.}  In addition to parameterize each user $a$ with a base embedding $\mathbf{p}_a$ and each item $i$ with a base embedding $\mathbf{w}_i$ as many latent factor based models~\cite{UAI2009bpr,KDD2008factorization}, we also take the inputs of the three social contextual aspects: $\mathbf{s}_a$, $\mathbf{l}_a$, and {\small$C_i$}. To model the complex contextual aspects, we extend the classical latent factor models and assume each user and each item has two embeddings. Specifically, each user $a$ is associated with a base embedding $\mathbf{p}_a$  from the base embedding matrix {\small $\mathbf{P}$} to denote her base latent interest in the standard latent factor based models, and an auxiliary  embedding vector $\mathbf{q}_a$ from the auxiliary embedding matrix {\small $\mathbf{Q}$} . This auxiliary user embedding vector characterizes each user's preference from the social contextual aspects that could not be detected by standard user-image rating behavior. Similarly, each image $i$ is also associated with two embeddings: a base embedding $\mathbf{w}_i$ from the item base embedding matrix {\small $\mathbf{W}$} to denote the basic image latent vector, and an auxiliary vector $\mathbf{x}_i$ from the item auxiliary embedding matrix {\small $\mathbf{X}$} to characterize each image from the social contextual inputs. Thus, by combining the attention mechanism with the embeddings, we model each user $a$'s predicted preference to image $i$ as a hierarchical attention:

\begin{small}
\vspace{-0.3cm}
\begin{flalign} \label{eq:pred_r}
& \hspace{12mm} \hat{r}_{ai}=\mathbf{w}^T_i{(\mathbf{p}_a + \gamma_{a1} \widetilde{\mathbf{x}}_a + \gamma_{a2} \widetilde{\mathbf{q}}_a +\gamma_{a3} \mathbf{q}_{C_i})} \nonumber \\
&\mbox{where}\quad  \widetilde{\mathbf{x}}_a= \sum_{j=1}^N l_{ja} \alpha_{aj}\mathbf{x}_j, \quad \widetilde{\mathbf{q}}_a=\sum_{b=1}^{M} s_{ba}\beta_{ab} \mathbf{q}_b.
\end{flalign}
\vspace{-0.3cm}
\end{small}

In the above prediction function, the representations of three contextual aspects are seamlessly incorporated in a holistic way. Specifically, the first line of Eq.\eqref{eq:pred_r} is a top layer attention network that aggregates the three contextual aspects for user embedding. The detailed attention subnetworks of the upload history attention and the social influence attention are listed in the second row. In fact, the attentive weights~($\gamma_{al}, \alpha_{aj}$, and $\beta_{ab}$) rely on our carefully designed attention networks that take various information as input. We leave the details of how to model these three attention networks in the following subsections. Next, we show the soundness of the objective predicted function.

\textbf{Relations to Other Models.} By rewriting the predicted preference score in Eq~\eqref{eq:pred_r}, we have:

\begin{small}
\vspace{-0.1cm}
\begin{flalign} \label{eq:pred_r_detail}
&\hat{r}_{ai}= \overbrace{\mathbf{p}_a^T\mathbf{w}_i}^{\mbox{Basic Latent Factor Model}}+ \underbrace{\gamma_{a1}\sum_{j=1}^N \alpha_{aj}l_{ja} \mathbf{x}_j^T\mathbf{w}_i}^{\mbox{Item Neighborhood Model}} \nonumber \\
&+\underbrace{\gamma_{a2} \sum_{b=1}^M s_{ba}\beta_{ab}\mathbf{q}^T_b \mathbf{w}_i}_{\mbox{Social Neighborhood Model}}+ \underbrace{\gamma_{a3} \mathbf{q}_{C_i}^T\mathbf{w}_i}_{\mbox{Owner Admiration Bias}},
\end{flalign}
\vspace{-0.2cm}
\end{small}

\noindent where the first part is a basic latent factor model, and the following three parts are extracted from the three contextual aspects. In the last three terms, $\mathbf{x}_j^T\mathbf{w}_i$ can be seen as the similarity function between image $i$ and the user's uploaded image $j$ in the neighborhood-based collaborative filtering from the upload history aspect~\cite{KDD2008factorization}. {\small$\mathbf{q}_b^T\mathbf{w}_i$} represents the social neighbor's preference to image $i$ with the social influence aspect. 
As each image is uploaded by a creator, the last term models the creator admiration aspect. This is quite natural in the real-world, as we always like to follow some specific creators' updates.

Please note that, if we replace all the attention scores with equal weights~(i.e., $\alpha_{ai}=\frac{1}{\sum_{j=1}^N l_{ja}},\beta_{ab}=\frac{1}{\sum_{b=1^M} s_{ba}}$, and $\gamma_{al}=\frac{1}{3}$), our model turns to an enhanced SVD++ model with rich social contextual information modeling~\cite{KDD2008factorization,SIGIR2017autosvd++}. However, this fixed weight assignment treats each user, each aspect, and the elements in each aspect equally. This simply configuration neglects that each user has different considerations for these three contextual aspects. By using hierarchical attention networks, we could learn each user's attentive weights from their historical behaviors.

\vspace{-0.1cm}
\subsection{Hierarchical Attention Network Modeling}

In this subsection, we would follow the bottom-up step to model the hierarchical attention networks in detail. Specifically, we would first introduce the two bottom layered attention networks: the upload history attention network and the social influence attention network, followed by the top layered aspect importance attention network that is based on the bottom layered attention networks.

\textbf{Upload History Attention.} The goal of the upload history attention is to select the images from each user $a$'s upload history that are representative to $a$'s preferences, and then aggregate this upload history contextual information to characterize each user. Given each image $j$ that is uploaded by $a$, we model the upload history attentive score $\alpha_{aj}$ as a three-layered attention neural network:

\begin{small}
\vspace{-0.2cm}
\begin{flalign} \label{eq:att_upload}
&\alpha_{aj}=\mathbf{w}^1 \times \sigma(\mathbf{W}^1 [\mathbf{p}_a, \mathbf{q}_a, \mathbf{x}_j, \mathbf{w}_j, \mathbf{e}_a,  \nonumber\\
&\mathbf{W}^c\mathbf{f}^c_j, \mathbf{W}^s\mathbf{f}^s_j, \mathbf{W}^c\mathbf{f}^c_a, \mathbf{W}^s\mathbf{f}^s_a])
\end{flalign}
\vspace{-0.2cm}
\end{small}

\noindent where {\small$\Theta_u=[ \mathbf{W}^c, \mathbf{W}^s, \mathbf{W}^1, \mathbf{w}^1]$} is the parameter set in this three layered attention network, and $\sigma(x)$ is a non-linear activation function. Specifically,  as the dimensions of visual content embeddings~(i.e., $\mathbf{f}^c_j$ and $\mathbf{f}^c_a$) and the visual style embeddings ~(i.e., $\mathbf{f}^s_j$ and $\mathbf{f}^s_a$) are much higher than the dimensions of other kinds of embeddings, {\small$\mathbf{W}^c\in\mathbb{R}^{D\times 4096}$} and {\small $\mathbf{W}^s \in \mathbb{R}^{D\times 5120}$} are the parameters of the bottom layer that performs dimension reduction of the visual content and style representations. {\small $\mathbf{W}^1\in \mathbb{R}^{(8D+d)\times d1}$} denotes the matrix parameter of the second layer in the attention network, as all data embedding vectors has {\small $D$} dimensions except the social embedding $\mathbf{e}_a$ has $d1$ dimensions. And $\mathbf{w}^1\in\mathbb{R}^{d1}$ is the vector parameter of the third layer in the attention network. In this attention modeling process, we take three different kinds of  embeddings as input:

\begin{itemize}
  \item \textbf{Latent Embedding}: the latent embedding includes $[\mathbf{p}_a, \mathbf{q}_a, \mathbf{x}_j, \mathbf{w}_j]$, where $\mathbf{p}_a$ and $\mathbf{q}_a$ are the basic and auxiliary embeddings of user $a$, and $\mathbf{x}_j$ and $\mathbf{w}_j$ are the basic and auxiliary embeddings of item $j$.
  \item \textbf{Social Embedding}: the social embedding part contains the learned social embedding $\mathbf{e}_a$ of each user, which models the global and local structure of each user in the social network $\mathbf{S}$.
  \item \textbf{Visual Embedding}: the visual embedding part includes the visual representations of user $a$ and item $j$. Specifically, each image is characterized by content representation $\mathbf{f}^c_j$ and style representation $\mathbf{f}^s_j$. Besides, as users show their preferences for images from their historical implicit feedbacks, each user $a$'s visual content representation and style representation can also be summarized as:  $\mathbf{f}^c_{a}=\frac{\sum_{i=1}^N r_{ai}\mathbf{f}^c_i}{\sum_{i=1}^N r_{ai}}, \mathbf{f}^s_{a}=\frac{\sum_{i=1}^N r_{ai}\mathbf{f}^s_i}{\sum_{i=1}^N r_{ai}}$.

\end{itemize}

%
%
%
%


By feeding all the sophisticated designed embeddings from heterogeneous data sources as the input, the upload history attention network learns to focus on the specific information. Please note that, we omit the bias terms in the attention network without confusion. In the following of this paper,  for ease of explanation, we also omit the dimension reduction for the visual embeddings~(i.e., {\small $\mathbf{W}^c$} and {\small $\mathbf{W}^s$}) whenever they are appeared for the attention modeling. Then, the final attentive upload history score $\alpha_{aj}$ is obtained by normalizing the above attention scores as:

\begin{small}
\vspace{-0.2cm}
\begin{equation} \label{eq:att_upload_nor}
\alpha_{aj}=\frac{exp(\alpha_{aj})}{\sum_{k=1}^N  exp(l_{ka}\alpha_{ak})}.
\end{equation}
\vspace{-0.2cm}
\end{small}

After we obtain the attentive upload history score $\alpha_{aj}$, the upload history context of user $a$, denoted as {\small$\widetilde{x}_a$}, is calculated as a weighted combination of the
learned attentive upload history scores:

\begin{small}
\vspace{-0.2cm}
\begin{equation} \label{eq:att_upload_c}
\widetilde{x}_a= \sum_{j=1}^N  l_{ja}\alpha_{aj}\mathbf{x}_j.
\end{equation}
\vspace{-0.2cm}
\end{small}

\textbf{Social Influence Attention.} The social influence attention module tries to select the influential social neighbors from each user $a$'s social connections, and then summarizes these social neighbors' influences into a social contextual vector. If user $a$ follows $b$, we use $\beta_{ab}$ to denote the social influence strength of $b$ to $a$. Then, the social attentive score $\beta_{ab}$ could be calculated as:

\begin{small}
\vspace{-0.2cm}
\begin{equation} \label{eq:att_social}
\beta_{ab}=\mathbf{w}^2\sigma(\mathbf{W}^2[\mathbf{p}_a, \mathbf{p}_b, \mathbf{q}_a, \mathbf{q}_b, \mathbf{e}_a, \mathbf{e}_b, \mathbf{f}^c_a, \mathbf{f}^s_a]),
\end{equation}
\vspace{-0.2cm}
\end{small}

\noindent where {\small$\Theta_s=[\mathbf{W}^2, \mathbf{w}^2]$} are the parameters in the social influence attention network. This social influence attention part also contains three kinds of data embeddings: the user interest embeddings of $\mathbf{p}_a, \mathbf{p}_b, \mathbf{q}_a, \mathbf{q}_b$, the social embeddings of $\mathbf{e}_a$ and $\mathbf{e}_b$, and the visual embeddings of user $a$ with content representation $\mathbf{f}^c_a$ and style representation $\mathbf{f}^s_a$.

Then, the final attentive social influence score $\beta_{ab}$ is obtained by normalizing the above attention scores as:

\begin{small}
\vspace{-0.2cm}
\begin{equation} \label{eq:att_social_nor}
\beta_{ab}=\frac{exp(\beta_{ab})}{\sum_{c=1}^M exp(s_{ca}\beta_{ac})}.
\end{equation}
\vspace{-0.2cm}
\end{small}

After we obtain the attentive social influence score $\beta_{ab}$, the social context of user $a$, denoted as {\small$\widetilde{q}_a$}, is calculated as the a weighted combination as:

\begin{small}
\vspace{-0.2cm}
\begin{equation} \label{eq:att_social_c}
\widetilde{q}_a= \sum_{b=1}^M s_{ba}\beta_{ab}\mathbf{q}_b.
\end{equation}
\vspace{-0.2cm}
\end{small}

Since each image is uploaded by one creator, for each image $i$, the corresponding uploader is represented as $C_i$. Correspondingly, the owner appreciation context could be simply represented as the the auxiliary embedding  $q_{C_i}$ from the user auxiliary embedding matrix {\small $\mathbf{Q}$}.

\textbf{Aspect Importance Attention Network.}
The aspect importance attention network takes the contextual representation of each aspect from the bottom layered attention networks as input,  and models the importance of each aspect in the user's decision process.  Specifically, for each pair of user $a$ and image $i$, we have two contextual representations from the bottom layer of HASC as: upload history contextual representation $\widetilde{\mathbf{x}}_a$, the social influence contextual representation $\widetilde{\mathbf{q}}_a$, and the owner appreciation contextual representation $\mathbf{q}_{C_i}$. Then, the aspect importance score $\gamma_{al}$~(l=1, 2, 3) is modeled with an aspect importance attention network as:

\begin{small}
\vspace{-0.2cm}
\begin{equation} \label{eq:att_fac}
\gamma_{al}=\mathbf{w}^3\sigma(\mathbf{W}^3 \mathbf{a}_l),
\end{equation}
\vspace{-0.2cm}
\end{small}

\noindent where {\small$\Theta_a=[\mathbf{W}^3, \mathbf{w}^3]$} is the parameter set of this attention network, and
$\mathbf{a}_l~(l=1,2,3)$ denotes the input of the top layered attention network, which is the output of the bottom layered attention networks, i.e., $\mathbf{a}_1=\widetilde{\mathbf{x}_a}$ is the upload history contextual representation, $\mathbf{a}_2=\widetilde{\mathbf{q}_a}$ is the social influence contextual representation, and
$\mathbf{a}_3=\mathbf{q}_a$ denotes the representation of current active user $a$.

Then, the final aspect importance score $\gamma_{al}$  is obtained by normalizing the above attention scores as:

\begin{small}
\vspace{-0.2cm}
\begin{equation} \label{eq:att_fac_nor}
\gamma_{al}=\frac{exp(\gamma_{al})}{\sum_{k=1}^3 exp(\gamma_{ak})}.
\end{equation}
\vspace{-0.2cm}
\end{small}

For each user $a$, the learned aspect importance scores are tailored to each user, which
distinguish the importance of the three social contextual aspects in the user's decision process.
For all learned aspect importance scores, the larger the value, the more likely the user's decision is
influenced by this corresponding social contextual aspect.

\subsection{Model Learning}
As we focus on implicit feedbacks of users, similar as the widely used ranking based loss function in ranking based latent factor models~\cite{UAI2009bpr}, we also design a ranking based loss function  as:

\begin{small}
\vspace{-0.2cm}
\begin{equation}\label{eq:loss_r}
\min\limits_{\Theta} \mathcal{L}=\sum_{a=1}^M\sum\limits_{(i,j)\in D_a }s(\hat{r}_{ai}-\hat{r}_{aj}) +\lambda||\Theta_1||^2
\end{equation}
\vspace{-0.2cm}
\end{small}

\noindent where  $s(x)$ is a sigmoid function that transforms the input into range $(0,1)$. {\small$\Theta\!=\![\Theta_1, \Theta_2]$}, with {\small$\Theta_1\!=\![\mathbf{P}, \mathbf{Q}, \mathbf{W}, \mathbf{X}]$} denotes the embedding matrices and  {\small$\Theta_2\!=\![\Theta_u,\Theta_s,\Theta_a]$} denotes the parameters in each attention network. $\lambda$ is a regularization term that regularizes the user and image embeddings. {\small$D_a=\{(i,j)|i\in R_a\!\wedge\!j\in V-R_a\}$} is the training data for $a$ with {\small$R_a$} the imageset that $a$ positively shows feedback.

All the parameters in the above loss function are differentiable. In practice, we implement HASC with TensorFlow to train model parameters with mini batch Adam. The detailed training algorithm is shown in Algorithm \ref{alg:sig}.
In practice, we could only observe positive feedbacks of users with huge missing unobserved values, similar as many implicit feedback works, for each positive feedback, we randomly sample 5 missing unobserved feedbacks as pseudo negative feedbacks at each iteration in the training process~\cite{TKDE2017modeling,WWW2017Visual,SIGIR2017attentive}.  As each iteration the pseudo negative samples change, each missing value gives very weak negative signal.
%
%

\begin{small}\vspace{-0.1cm}
{\renewcommand\baselinestretch{0.8}\selectfont
\begin{algorithm}[htb]
\renewcommand{\algorithmicrequire}{\textbf{Input:}}
\renewcommand\algorithmicensure {\textbf{Output:}}
\caption{ \small{The learning algorithm of HASC}\ \ \ \ \ \  }\label{alg:sig}
\begin{algorithmic}[1]
\REQUIRE Rating matrix {\small$\mathbf{R}$}, social matrix {\small$\mathbf{S}$}, Uploader matrix {\small $\mathbf{L}$}; batch size $m$; max epoch $T$;\\
\ENSURE Latent embedding matrix {\small $\Theta_1\!\!=\!\![\mathbf{P,Q,W,X}]$} and parameters in the attention networks $\Theta_2$ ;\\

\STATE Initialize {\small$\Theta$}  with a Gaussian distribution with a mean of 0 and a standard variation of 0.1;\\
\FOR {epoch $\gets 1$ to $T$}
    \STATE Get training data {\small$\mathcal{D}$} with randomly selected 5 times negative feedbacks：\mbox{$<a,i,j>$} ({\small$a\!\in\!U, i\!\in\!R_a, j\!\in\!V\!-\!R_a$});
    \FOR {mini epoch $\gets 1$ to $\frac{|\mathcal{D}|}{m}$}
        \STATE Get mini batch : randomly select m pairs \mbox{$<{a^k, i^k, j^k}>_{k=1}^m$} in the training data;
        \FOR {Each pair {\small $<a^k,i^k,j^k>$ } in the mini batch}
            \STATE Compute predicted rating of positive item~{\small $\small \hat{r}_{ai}$}~(Eq.\eqref{eq:pred_r});
            \STATE Compute predicted rating of negative item~{\small $\small \hat{r}_{aj}$}~(Eq.\eqref{eq:pred_r});
            \STATE Compute the loss $\mathcal{L}^k$ ~(Eq.\eqref{eq:loss_r});
        \ENDFOR
        \STATE Update $\Theta$ with loss as $\frac{1}{m}\sum_{k=1}^m\mathcal{L}^k$;
    \ENDFOR
\ENDFOR
\STATE Return {\small $\Theta_1=[\mathbf{P,Q,W,X}]$} and parameters in the attention $\Theta_2$.

\end{algorithmic}
\end{algorithm}
\par}
\end{small}
\vspace{-0.2cm}

\section{Experiments}
In this section, we show the effectiveness of our proposed HASC model. Specifically, we would answer the following questions: Q1: How does our proposed model perform compared to the baselines~(Sec.~\ref{sec:exp_ove})? Q2: How does the model perform under different sparsity~(Sec.~\ref{sec:exp_spasity})? Q3: How does the proposed social contextual aspects and the hierarchical attention perform~(Sec.~\ref{sec:exp_att})?


\begin{figure*} [htb]
  \begin{center}
  \vspace{-0.2cm}
      \subfigure[\vspace{-1.5cm}HR@K on F\_S]{\includegraphics[width=80mm]{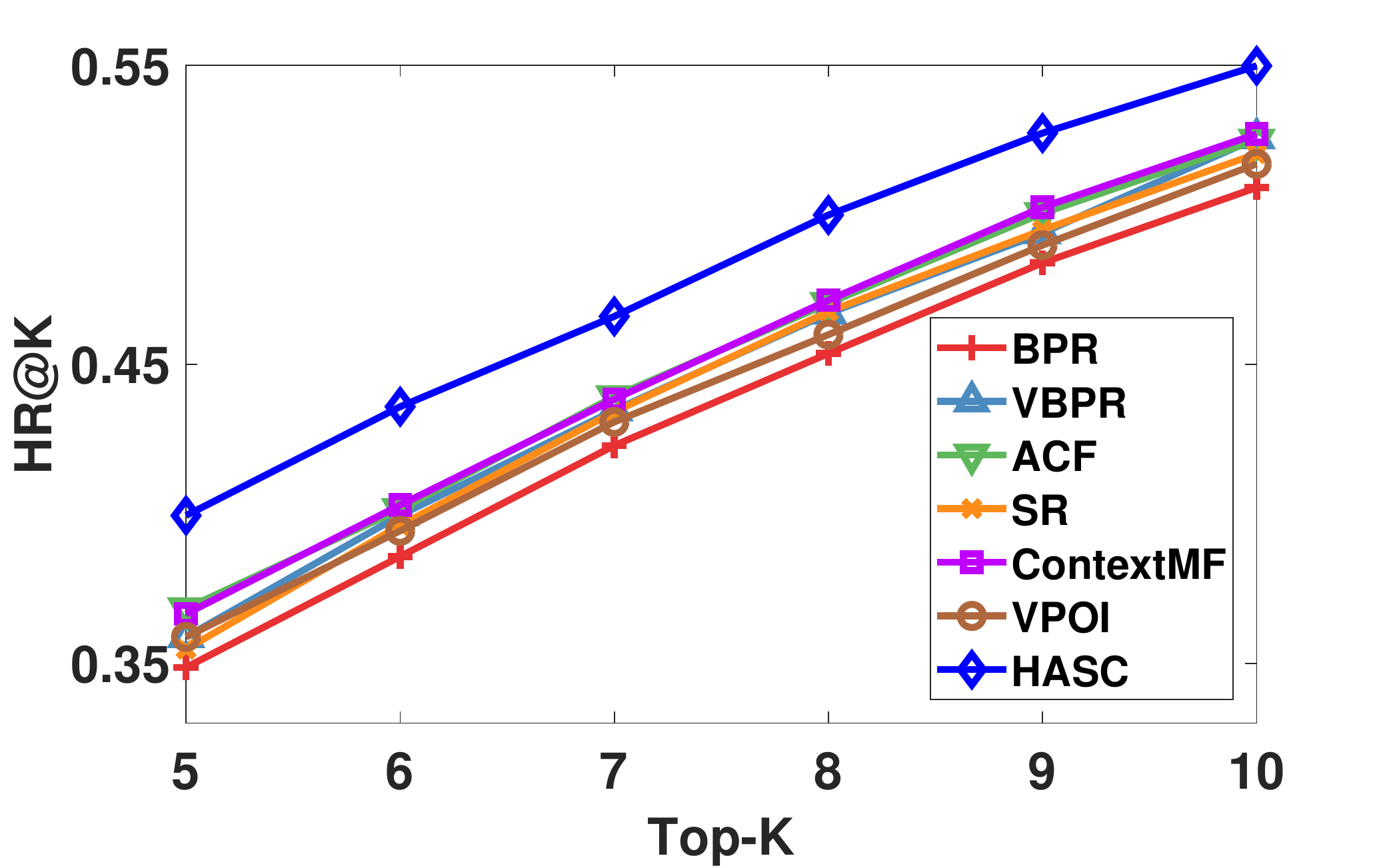} \label{fig:hit_s}}
      \subfigure[\vspace{-1.5cm}NDCG@K on F\_S]{\includegraphics[width=80mm]{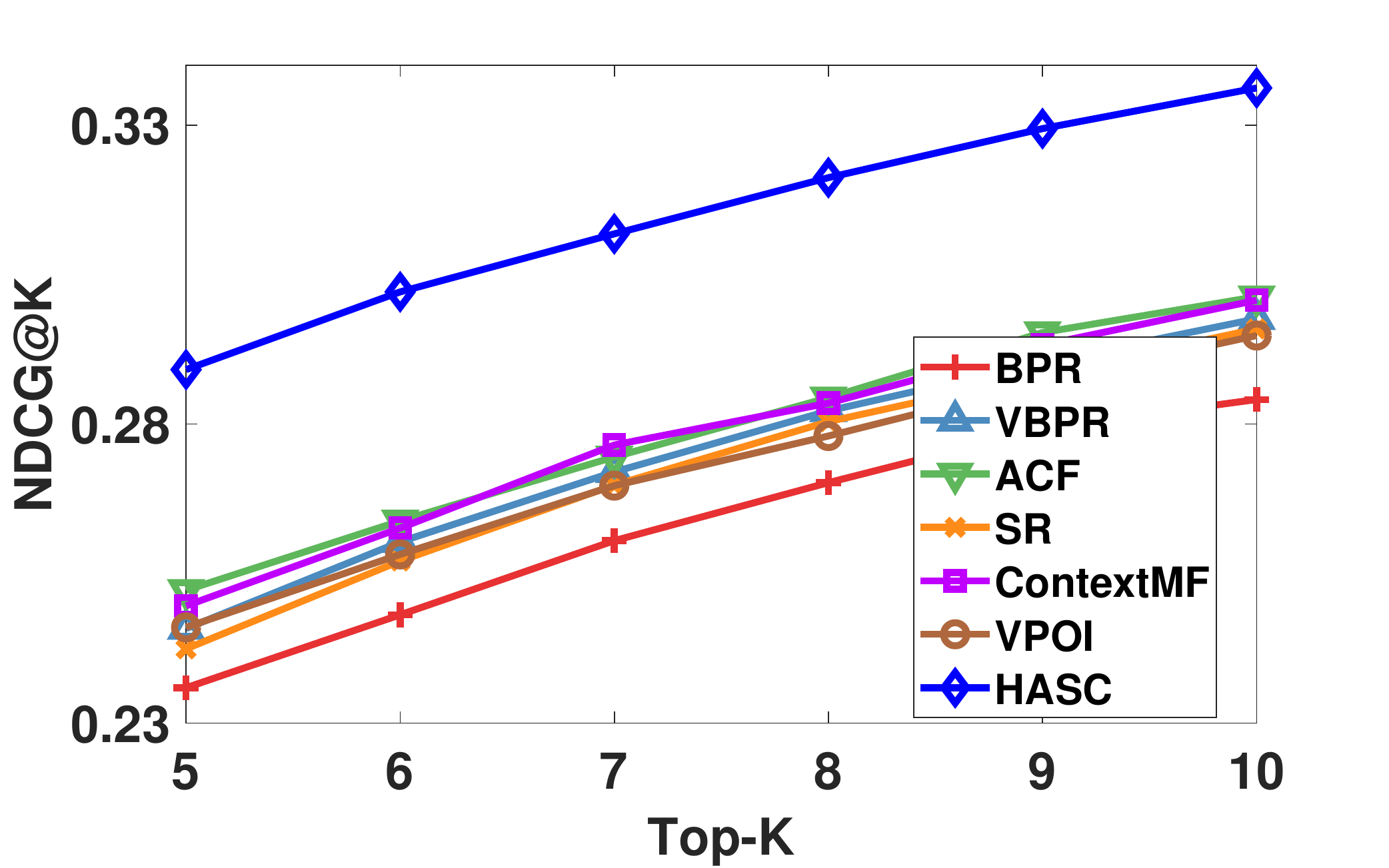}  \label{fig:ndcg_s}}
  \vspace{-0.2cm}
      \subfigure[\vspace{-1.5cm}HR@K on F\_L]{\includegraphics[width=80mm]{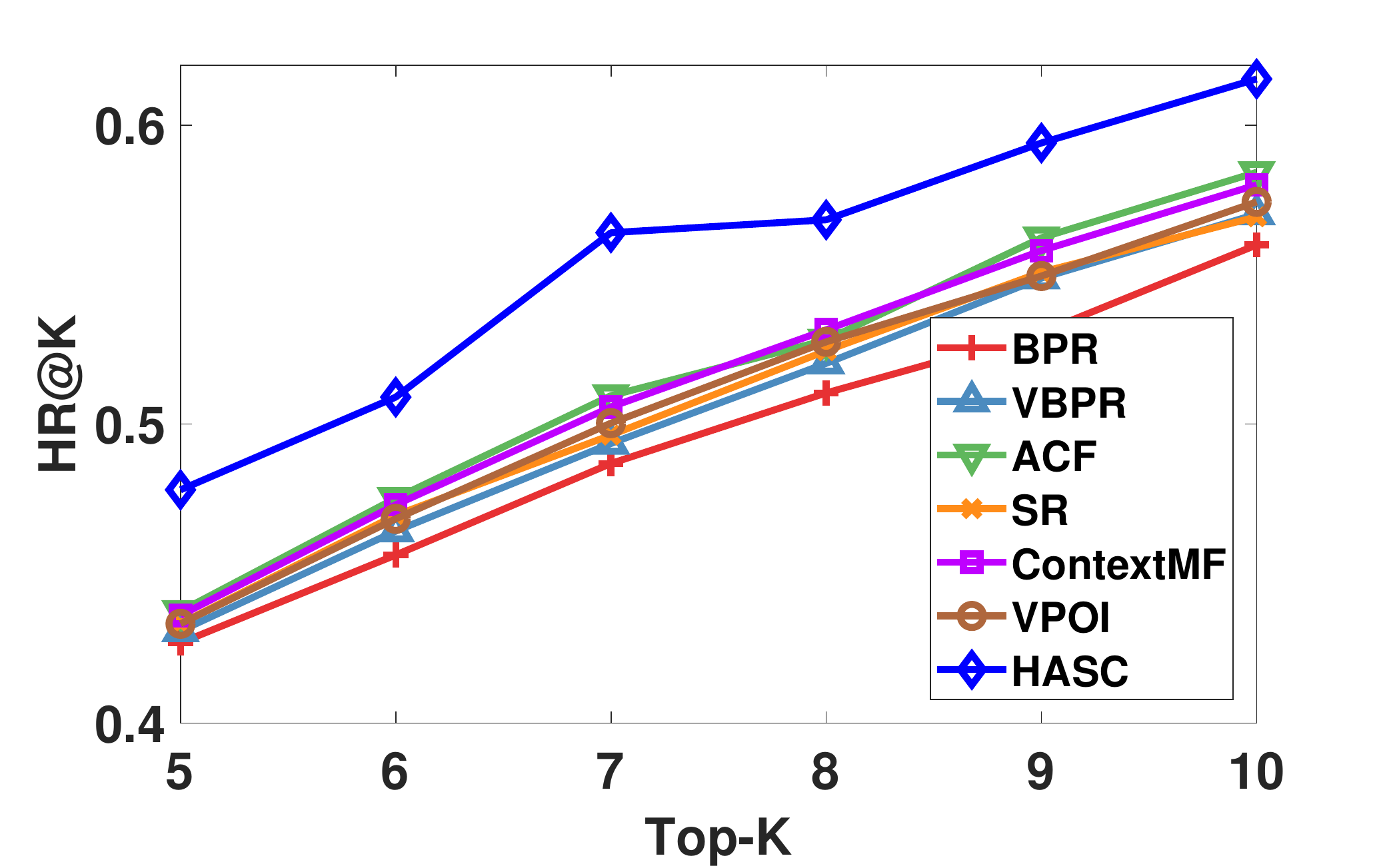} \label{fig:hit_l}}
     \subfigure[\vspace{-1.5cm}NDCG@K on F\_L]{\includegraphics[width=80mm]{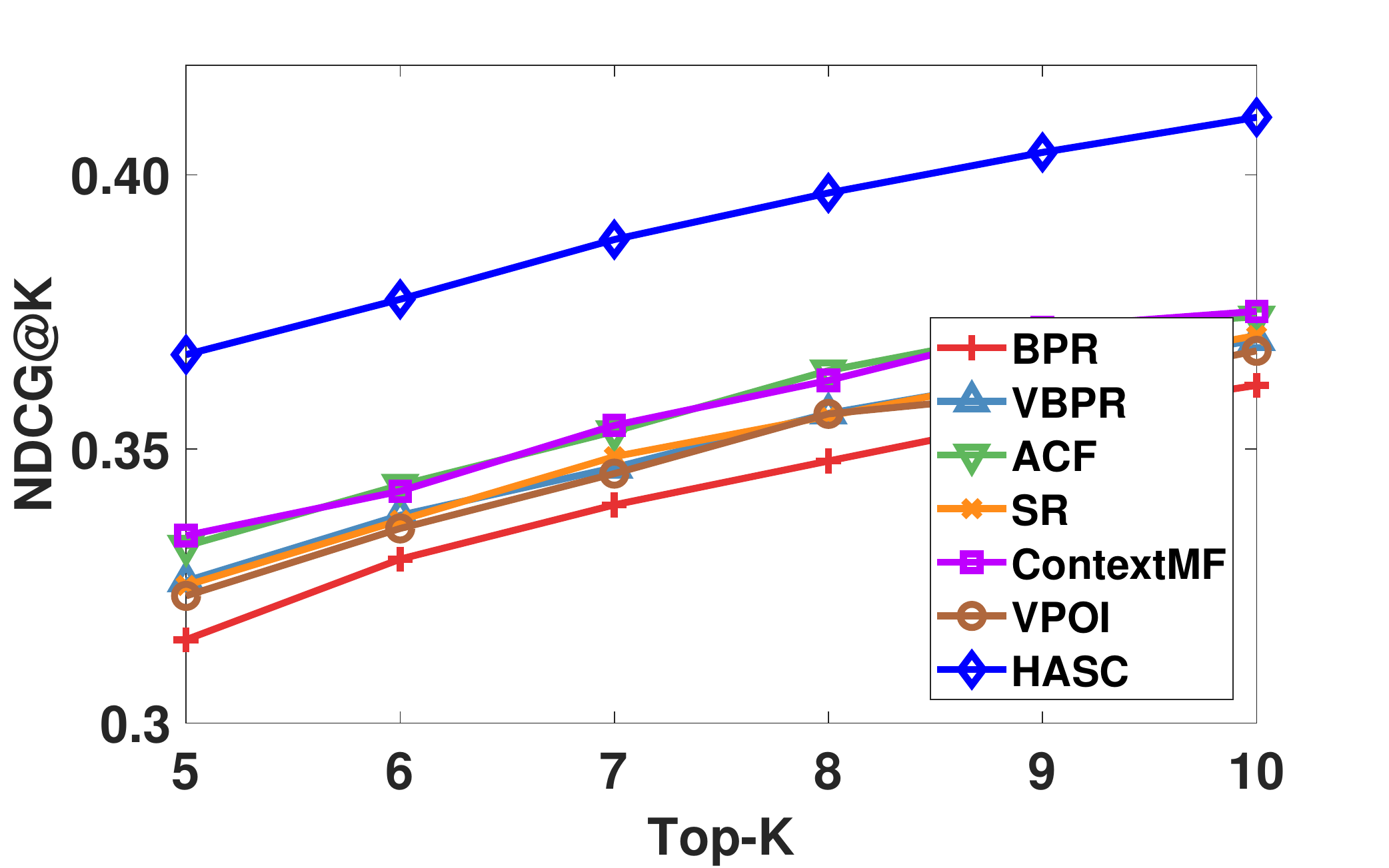}  \label{fig:ndcg_l}}
  \end{center}\vspace{-0.4cm}
  \caption{\small{Overall performance of different models on the two datasets. (Better viewed in color.)}} \label{fig:ov_rank}
  \vspace{-0.4cm}
\end{figure*}

\subsection{Experimental Settings}\label{sec:exp_set}

\textbf{Dataset.} To the best of our knowledge, there is no public available dataset that contains heterogenous data sources in a social image based network as described in Fig.~\ref{fig:data}.
To show the effectiveness of our proposed model, we crawl a large dataset from one of the largest social image sharing platform \emph{Flickr}, which is extended from the widely used NUS-WIDE dataset~\cite{MM2009nus,PAMI2017tri}. NUS-WIDE contains nearly 270,000 images with 81 human defined categories from Flickr. Based on this initial data, we get the uploader information according to the image IDs provided in NUS-WIDE dataset from the public APIs of Flickr. We treat all the uploaders as the initial userset, and the associated images as  the imageset. We then crawl the social network of the userset, and the implicit feedbacks of the userset to the imageset.

After data collection, in data preprocessing process, we filter out users that have less than 2 rating records and 2 social links. We also filter out images that have less than 2 records. We call the filtered dataset as \emph{F\_L}. As shown in Table~\ref{tab:data_stats}, this dataset is very sparse with about 0.15\% density. Besides, we further filter $F\_L$ dataset to ensure each user and each image have at least 10 rating records. This leads to a smaller but denser dataset as \emph{F\_S}. Table~\ref{tab:data_stats} shows the statistics of the two datasets after pruning. Please note that the number of images is much more than that of the users. This is consistent with the observation that the number of  images usually far exceeds that of users in social image platforms~\cite{wwwflickr}, as each user could be a creator to upload multiple images. In data splitting process, we follow the leave-one-out procedure in many research works~\cite{SIGIR2017attentive,WWW2017neural}. Specifically, for each user, we select the last rating record as the test data, and the remaining data are used as the training data. To tune model parameters, we randomly select 5\% of the training data to constitute the validation dataset.

\begin{table}[htb] \centering
\caption{\small{The statistics of the two datasets.}}\label{tab:data_stats}
\vspace{-0.3cm}
\begin{scriptsize}
\begin{tabular}{|l|c|c|c|c|c|}
\hline
Dataset & Users& Images &Ratings & Social Links & Rating Density \\ \hline
F\_S& 4,418 &31,460 & 761,812 &184,991 & 0.55\%  \\
F\_L&8,358 &105,648 & 1,323,963, & 378,713 &0.15\% \\ \hline
\end{tabular}
\end{scriptsize}
\vspace{-0.2cm}
\end{table}

\textbf{Evaluation Metrics} Since we focus on recommending images to users, we use two widely adopted ranking metric for \emph{top-K} recommendation evaluation: the Hit Ratio~(HR) and Normalized Discounted Cumulative Gain~(NDCG)~\cite{AAAI2016vbpr,SIGIR2017attentive}. HR measures the percentage of images that are liked by users in the top-K list, and NDCG gives a higher score to the hit images that are ranked higher in the ranking list.  As the image size is huge, it is inefficient to take all images as candidates to generate recommendations. For each user, we randomly select 100 unrated images as candidates, and then mix them with the records in the validation and test data to select the top-K results. This evaluation process is repeated for 10 times and we report the average results~\cite{AAAI2016vbpr,SIGIR2017attentive}. For both metrics, the larger the value, the better the ranking performance.

\textbf{Baselines.}
We compare our proposed HASC model with the following baselines:

\vspace{-0.1cm}
\begin{itemize}
   \item \emph{BPR}: it is a classical ranking based latent factor based model for recommendation with competing performance. This method has been well recognized as a strong baseline for recommendation~\cite{UAI2009bpr}.
  \item \emph{SR}: it is a social based recommendation model that encodes the social influence among users with social regularization in classical latent factor based models~\cite{WSDM2011recommender}.
  \item \emph{ContextMF}: this method models various social contextual factors, including item content topic, user personal interest, and inter-personal influence in a unified social contextual recommendation framework~\cite{TKDE2014scalable}.
  \item \emph{VBPR}: it extends BPR by modeling both the visual and latent dimensions of users' preferences in a unified framework, where the visual content dimension is derived from a pre-trained VGG network.
  \item \emph{ACF}: it models the item level and component level attention for image recommendation with two attention networks. For fair comparison, we enrich this baseline by leveraging the upload history as users' auxiliary feedback in this model~\cite{SIGIR2017attentive}.
  \item \emph{VPOI}: it is a visual based POI recommendation algorithm. This algorithm relies on the collective matrix factorization to consider the associated images with each POI and the uploaded images of each user.
      To adapt the POI recommendation to image recommendation, we treat each image as a POI and the uploaded images
     of each user as the associated images of her.~\cite{WWW2017Visual}.
\end{itemize}

\vspace{-0.1cm}


\textbf{Parameter setting.} In the social embedding process with Deepwalk~\cite{KDD2014deepwalk}, we set the parameters as:  the window size $w=10$ and walks per vertex $\rho=80$. The social embedding size $d$ is set in the range $[32,64,128]$. We find when $d=128$, the social embedding reaches the best performance. Hence, we set $d\!=\!128$ in Deepwalk. There are two important parameters in our proposed model: the dimension {\small$D$} of the user and image embeddings, and the regularization parameter $\lambda$ in the objective function~(Eq.\eqref{eq:loss_r}). We choose {\small$D$}  in {\small$[10,15,20,30]$} and $\lambda$ in $[0.001,0.01,01]$, and perform grid search to find the best parameters. The best setting is {\small$D\!=\!15$} and $\lambda=0.01$. We find the dimension of the attention networks does not impact the results much. Thus, we empirically set the dimensions of the parameters in the attention networks as 20~(i.e., parameters in $\Theta_2$). The activation function $\sigma(x)$ is set as the Leakly ReLU. To initialize the model, we randomly set the weights in the attention networks with a Gaussian distribution of mean 0 and standard deviation 0.1. Since the objective function of HASC is non-convex, we initialize {$\mathbf{P}$} and {$\mathbf{W}$} from the basic BPR model, and {$\mathbf{Q}$} and {$\mathbf{X}$} with the same Gaussian distribution as the parameters of the attention networks to speed up convergence. We use mini-batch Adam to optimize the model, where the batch size is set as 512 and the initial learning rate is set as 0.0005. There are several parameters in the baselines, for fair comparison, all the parameters in the baselines are also tuned to have the best performance. For all models, we stop model training when both the HR@5 and NDCG@5 on the validation dataset begins to decrease.

\begin{figure*} [htb]
  \begin{center}
  \vspace{-0.2cm}
  \includegraphics[width=140mm]{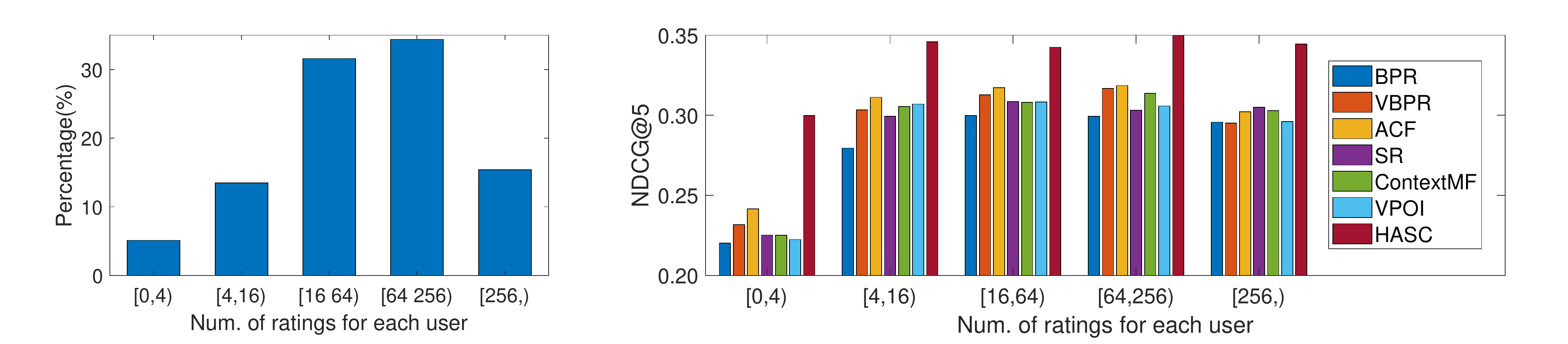}
  \end{center}\vspace{-0.4cm}
  \caption{\small{Performance under different sparsity.}} \label{fig:bin_results}
  \vspace{-0.2cm}
\end{figure*}

\vspace{-0.1cm}
\subsection{Overall Performance} \label{sec:exp_ove}
Fig.~\ref{fig:ov_rank}  shows the overall performance of all models on HR@K and NDCG@K on the two datasets with varying sizes of $K$, where the top two subfigures depict the results on F\_S dataset and the bottom two subfigures depict the results on F\_L dataset. As shown in this figure, our proposed HASC model always performs the best. With the increase of the top-K list size, the performance of all models increase. The performance trend is consistent over different top-K values and different metrics. We find that considering either the social network or the visual image information could alleviate the data sparsity problem and improve recommendation performance. E.g., VBPR improves over BPR about 3\% by incorporating the visual information in the modeling process. ACF further improves VBPR by assigning the attentive weights to different images the user rated and uploaded in the past. SR also has better performance as it leverages the social network information, and ContextMF further improves the performance with content modeling. On average, our proposed model shows  about 20\%  improvement over BPR baseline, and more than 10\% improvement over the best baselines on both datasets with regard to the NDCG@5 metric. Last but not the least, by comparing the results of F\_S and F\_L, we observe that for each method, the results on F\_L always outperform F\_S. We guess a possible reason is that, though F\_S is denser than F\_L, the larger F\_L has nearly two times as many records as F\_S for training. As the overall trend is similar on the two metrics with different values of $K$, in the following of the subsections, for page limit, we only show the top-5 results.


\vspace{-0.1cm}
\subsection{Performance under Different Data Sparsity} \label{sec:exp_spasity}
A key characteristic of our proposed model is that it alleviates the data sparsity issue with various social contextual aspects modeling. In this subsection, we investigate the performance of various models under different data sparsity. We mainly focus on the F\_L dataset as it is more challenging with sparser user rating records compared to the denser F\_S dataset. Specifically, we bin users into different groups based on the number of the observed feedbacks in the training data, and then show the performance under different groups.  Fig.~\ref{fig:bin_results} shows the results, where the left part summarizes the user group distribution of the training data and the right part depicts the performance with different data sparsity. As shown in the left part, more than 5\% users have less than 4 ratings, and 20\% users have less than 16 ratings with more than  100 thousand images on the F\_L dataset.  When the rating scale is very sparse, the BPR baseline can not work well under this situation as it only modeled the  sparse user-image implicit feedbacks. Under this situation, the improvement is significant for all models over BPR as these models utilized different auxiliary data for recommendation. E.g., when users have less than 4 ratings, our proposed HASC model improves over BPR by more than 35\%.  As user rating scale increases, the performance of all models increase quickly with more training rating records, and HASC still consistently outperforms the baselines.


\begin{small}
\begin{table}[!htbp]
\centering
\vspace{-0.2cm}
\caption{\small{The improvement of using different attention mechanism compared to BPR.}
}\label{tab:tabatt_a}
\vspace{-0.2cm}
\begin{tabular}{|*{6}{c|}}
\hline
Bottom Layer &Top Layer&\multicolumn{2}{|c|}{F\_S}&\multicolumn{2}{|c|}{F\_L}\\
\cline{3-6}
Attention& Attention & HR&NDCG&HR&NDCG\\ \hline
AVG &AVG& 6.44\%& 10.28\% &5.54\% &9.02\%\\ \hline
MAX&MAX          &5.82\% &9.55\%  &4.98\%  &8.10\% \\ \hline
AVG&ATT          &7.33\% &11.15\% &5.95\% &9.93\%  \\  \hline
MAX&ATT         &6.84\%  &10.96\% &5.72\% & 9.55\% \\  \hline
ATT&AVG        &12.75\%  &19.23\%  &8.30\% &13.28\%  \\  \hline
ATT&MAX     &12.20\%   &18.56\%   &8.02\%  &12.85\% \\ \hline
ATT&ATT    &\textbf{14.57\%}&\textbf{22.55\%}&\textbf{10.67\%}&\textbf{16.70\%}\\
\hline
\end{tabular}
\vspace{-0.2cm}
\end{table}
\vspace{-0.2cm}
\end{small}

\begin{small}
\begin{table}[!htbp]
\centering
\vspace{-0.2cm}
\caption{\small{The improvement of modeling different contextual aspects with our proposed model compared to BPR.(U:upload history, S: social influence, C: creator admiration) }
}\label{tab:tabatt_b}
\vspace{-0.2cm}
\begin{tabular}{|*{5}{c|}}
\hline
\multirow{2}{*}{Aspects}&\multicolumn{2}{|c|}{F\_S}&\multicolumn{2}{|c|}{F\_L}\\
\cline{2-5}
 & HR&NDCG&HR&NDCG\\ \hline
U    & 8.70\% &16.52\%  &6.44\%  &11.03\% \\ \hline
S          &9.63\%  &16.78\% & 5.29\% & 9.65\% \\  \hline
C      &8.57\%  & 14.53\%  & 4.37\% &7.93\%  \\  \hline
U+S+C    &\textbf{14.57\%}&\textbf{22.55\%}&\textbf{10.67\%}&\textbf{16.70\%}\\
\hline
\end{tabular}
\vspace{-0.2cm}
\end{table}
\vspace{-0.2cm}
\end{small}

\subsection{Attention Analysis} \label{sec:exp_att}
In this part, we conduct experiments to give more detailed analysis of the proposed attention network. We would evaluate the soundness of the designed attention structure and the superiority of combining the various data embeddings for attention modeling.

In the experiments, we use the Leakly ReLU as the activation function $\sigma(x)$ for attention modeling, and then
attentively combine the elements of each set with a soft attention. Alternately, instead of attentively combining all the elements, a direct solution is to use the hard attention with MAX operation that selects the element with the largest attentive score at each layer of the hierarchical attention network. E.g., for the upload history aspect, \emph{Max} learns the attentive upload history score in Eq.\eqref{eq:att_upload_c} as: $\widetilde{\mathbf{x}}_a=\mathbf{x}_j,\quad\text{where} \quad l_{ja}\!=\!1\wedge (\forall l_{ka}=1,\alpha_{ja}\geq \alpha_{ka})$. Particularly, if we simply set the attentive scores with the average pooling~(i.e., $\alpha_{ai}=\frac{1}{|L_a|}, \beta_{ab}=\frac{1}{|S_a|}, \gamma_{al}=\frac{1}{3}$), our model degenerates to an enhanced SVD++ with social contextual modeling but without any attentive modeling. If we do not model any social contextual aspects, our model degenerates to the BPR model~\cite{UAI2009bpr}. Table~\ref{tab:tabatt_a} shows the results of different attention mechanism. As shown in this table, the best results are achieved by using our proposed attention mechanism, followed by AVG and MAX. We guess a possible reason is that: each user's interests are diversified, and it is challenging to infer each user's interests from the limited training data. If we simply using a hard attention with the maximum value or adopting average aggregation, many valuable contextual information is neglected in this process. Besides, we observe that ATT that operates at the bottom layer achieves much better performance than its counterparts
that operates on the top layer~(e.g., the comparison results between the fourth row and the sixth row). Since each aspect at the bottom layer usually contains much more elements than the top layer, attentively summarizing each contextual aspect at the bottom layer would provide valuable information for the top layer. In contrast, if we use
AVG or MAX at the bottom layer, the results are not satisfactory when we use ``ATT'' at the second layer, since the input of the second layer lacks many important information.

After showing the soundness of our proposed attention structure, Table~\ref{tab:tabatt_b} presents the performance of using different contextual aspects with our proposed hierarchical attention.  As shown in this table, each aspect improves the performance. By combining all social contextual aspects with hierarchical attention, the model reaches the best performance.

\begin{small}
\begin{table}[!htbp]
\centering
\caption{\small{Performance of different kinds of  inputs for attention modeling
(Base: base embedding, Aux: auxiliary embedding, Soc: social embedding, Vis\_C(Vis\_C): visual content(style) embedding). with ``Base'' denotes the base embedding, ``Aux'' denotes the auxiliary embedding, "Soc" denotes the social embedding,  and ``Vis\_C'', ``Vis\_S", ``Vis\_{CS}" denotes the visual content feature, visual style feature, and both visual features.
}}\label{tab:att_f}
\vspace{-0.2cm}
\begin{tabular}{|*{5}{c|}}
\hline
\multirow{2}{*}{Input Embedding}&\multicolumn{2}{|c|}{F\_S}&\multicolumn{2}{|c|}{F\_L}\\
\cline{2-5}
&HR&NDCG&HR&NDCG\\
\hline
Base& 0.358& 0.257& 0.439&0.319\\
\hline
Base+Aux       & 0.366 &0.264  &0.445  & 0.323 \\
\hline
Base+Aux+Soc     & 0.367 &0.270  &0.450  & 0.331 \\
\hline
Base+Aux+Vis\_C   &0.388 &0.278 &0.453 &0.335  \\
\hline
Base+Aux+Vis\_S  & 0.383 &0.275  &0.451  &0.332  \\
\hline
Base+Aux+Vis\_{CS}  &0.393 &0.282  &0.464  & 0.342  \\
\hline
Base+Aux+Soc+Vis\_{CS} &\textbf{0.400} &\textbf{0.289}  &\textbf{0.475}  &\textbf{0.347}  \\
\hline
\end{tabular}
\end{table}
\end{small}

\begin{small}
\begin{table}[!htbp]
\centering
\vspace{-0.2cm}
\caption{\small{Performance of different kinds of social embedding techniques for the attention modeling. }
}\label{tab:tabatt_network}
\vspace{-0.2cm}
\begin{tabular}{|*{5}{c|}}
\hline
\multirow{2}{*}{Input Embedding}&\multicolumn{2}{|c|}{F\_S}&\multicolumn{2}{|c|}{F\_L}\\
\cline{2-5}
&HR&NDCG&HR&NDCG\\
\hline
Base+Aux+DeepWalk     & 0.367 &0.270  &0.450  & 0.331 \\
\hline
Base+Aux+LINE   &    0.369  &0. 273             &0.452  &0.334  \\
\hline
Base+Aux+GCN    &    0.371  &0.276              &0.459  &0.340 \\
\hline
Base+Aux+DeepWalk+Vis\_{CS}  &0.400 &0.289  &0.475  &0.347  \\
\hline
Base+Aux+LINE+Vis\_{CS} &0.400 &0.289  &0.474  &0.345\\
\hline
Base+Aux+GCN+Vis\_{CS} &0.401 &0.290  &0.475 &0.348  \\
\hline
\end{tabular}
\vspace{-0.2cm}
\end{table}
\end{small}

Besides, in the attention modeling process, we also learn the attentive weights by modeling different kinds of input embeddings from the heterogeneous data sources. For each attention layer, it consists the following kinds of inputs: the latent interest representations of base embeddings~(i.e., $\mathbf{p}_a$ and $\mathbf{w}_i$) and auxiliary embeddings~(i.e., $\mathbf{q}_a$ and $\mathbf{x}_i$), the social embeddings~(i.e., $\mathbf{e}_a$), and the visual embeddings with content representations~(i.e., $\mathbf{f}^c_i$ of image $i$ and $\mathbf{f}^c_a$ of user $a$)  and style representations~(i.e., $\mathbf{f}^s_i$ of image $i$ and $\mathbf{f}^s_a$ of user $a$ ).  Table~\ref{tab:att_f} shows the  performance of HASC with different kinds of input embeddings.
From this table, we have several observations. First, as the auxiliary latent embedding representation could model each user and each item from the rich social contextual information, taking the auxiliary embeddings could improve the performance than solely feeding the base embeddings for attention modeling. Second, the improvement of social embeddings is not very significant. We guess a possible reason is that, the social influence aspect already considers the social neighborhood information for users' interest modeling.  As the social embeddings represent the overall social network with both local and global structure, the improvement is limited with the additional global network structure modeling. Third, we observe that the improvement of the visual embeddings is very significant. Both the content and the style information could enhance the recommendation performance. By combining content and style embeddings, the performance further improves. This observation empirically shows the complementary relationship of content and style in visual images. Last but not least, by feeding the three different kinds of data embeddings into the attention network embedding, the proposed HASC could achieve the best performance.

In the previous experiments, we use the DeepWalk as the social network embedding model to obtain the social network embedding  vector of each user. Now we would show the effectiveness of adopting different network embedding techniques. We choose two state-of-the-art network embedding models:  LINE~\cite{tang2015line} and GCN~\cite{ICLR2017semi}, and compare the performance. The results are shown in Table~\ref{tab:tabatt_network}. As can be seen from this table, when the item visual embeddings are not incorporated, using the advanced graph embedding techniques~(e.g., GCN), could partially improve the recommendation performance, as these advanced models could better capture the social network structure. When all the input embeddings are incorporated, these advanced graph embedding models show similar performance compared to the DeepWalk based network embedding model. We guess the reason is that, as stated as Table~\ref{tab:att_f}, the improvement of the social embedding is not as significant as the visual based input for attention modeling when all the input embeddings are considered.

\begin{figure} [htb]
  \begin{center}
  \vspace{-0.2cm}
  \includegraphics[width=80mm]{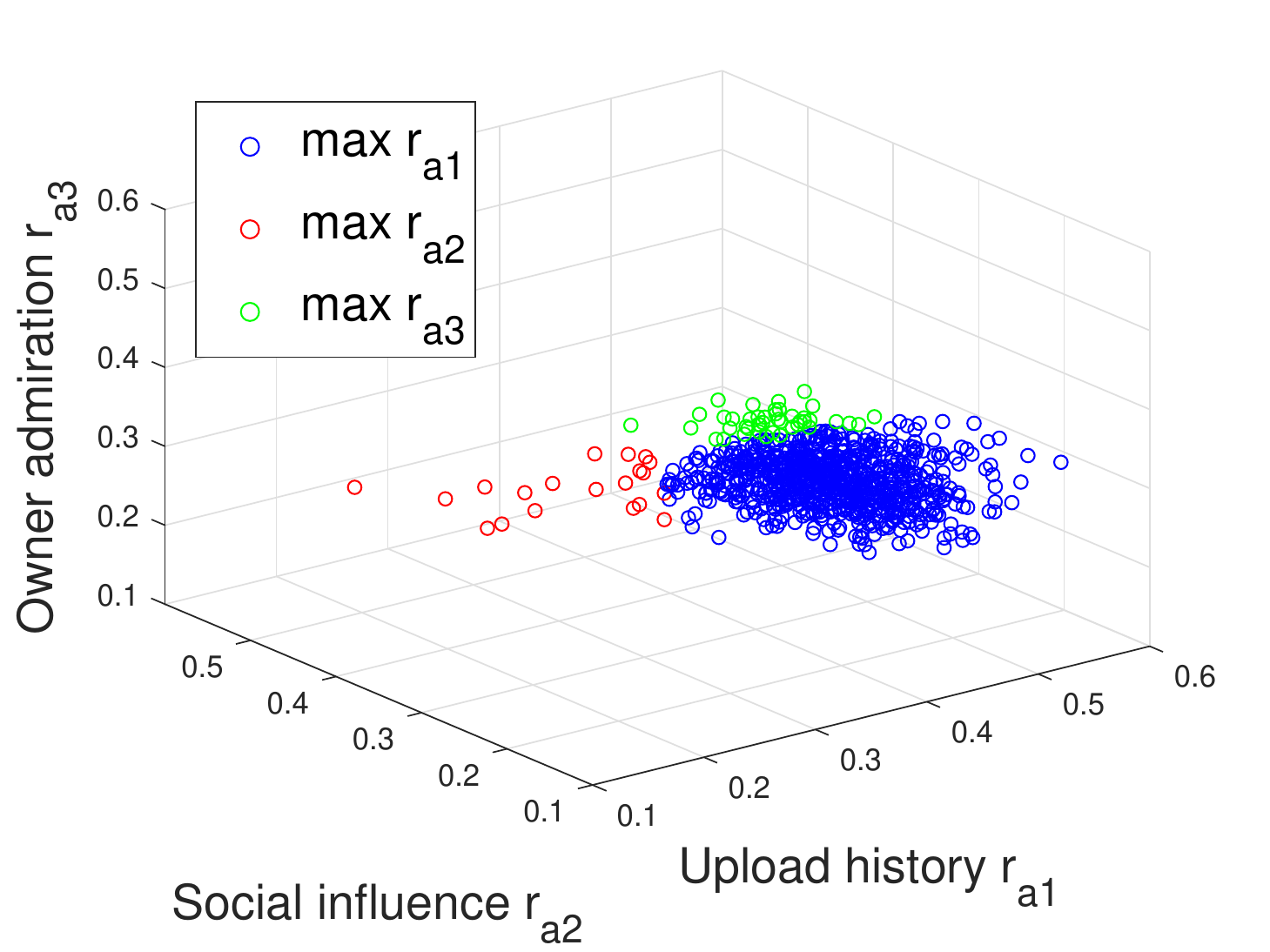}
  \end{center}\vspace{-0.6cm}
  \caption{\small{Visualization of aspect weights of randomly sampled users.}} \label{fig:att_vis}
  \vspace{-0.2cm}
\end{figure}

\textbf{Attention Weights Visualization.}
Besides given the overall results of different attention modeling setting, we give a visualization of the learned attention weights of users from the F\_L dataset. Firstly,  for each user, we group her into three categories according the aspect that has the largest attention value. In other words, for each user $a$ in the first group, she has the largest aspect weight for upload history, i.e., $\gamma_{a1}>\gamma_{a2}\land \gamma_{a1}>\gamma_{a3}$. Then, for each group, we randomly select 10\% of users and visualize them in Fig.~\ref{fig:att_vis}. As observed in this figure, each randomly sampled user has her own attentive weights for balancing the three contextual aspects. Besides, most of the users belong to the first group that has the largest value of the upload history aspect, which
empirically shows that many users show similar preferences between their uploaded images and the liked images. This observation is also consistent with Table~\ref{tab:tabatt_b} that shows
leveraging the upload history has the largest performance gain compared to the remaining two aspects on F\_L dataset.

\begin{figure*} [htb]
  \begin{center}
  \vspace{-0.2cm}
  \includegraphics[width=160mm]{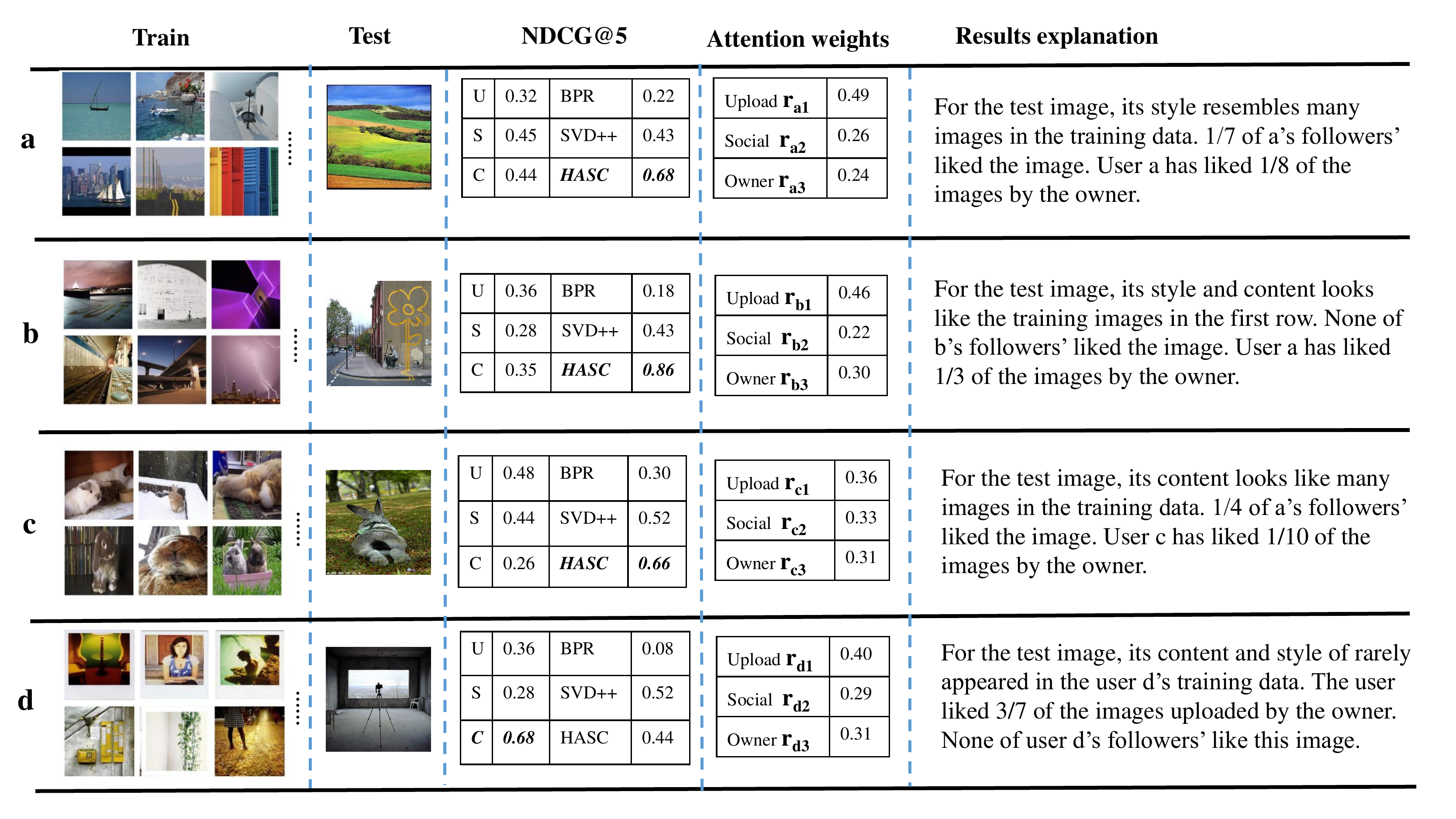}
  \end{center}\vspace{-0.6cm}
  \caption{\small{The case study of several typical users. In this figure, each row represents a user. The first and the second column are the training and test images of the user. The Top-5 recommendation results of NDCG@5 are shown in the third column. In the third column, the left three models are simplified versions of our proposed HASC model that only leverage one aspect, and the model with best performance is shown with bold italic letters.}} \label{fig:case}
  \vspace{-0.4cm}
\end{figure*}

\subsection{Case Study}  \label{sec:exp_vis}
In order to better understand the proposed model,  we visualize several typical users and the experimental results of different recommendation models in Fig.\ref{fig:case}. In this figure, each row represents a user. The first column shows the images liked by the user in the training data, and the second column shows the test image of each user in the test data.  Please note that, due to page limit, we only show six typical training images of each user if she has rated more than 6 images in the training data. The third column shows the NDCG@5 results of different models. Specifically, to validate the effectiveness of different aspects in the modeling process, we use \emph{U},  \emph{S}, and \emph{C} to denote the three simplified versions of our proposed HASC model that only consider the upload history aspect~(i.e., $\gamma_{a2}=\gamma_{a3}=0$), the social influence aspect(i.e., $\gamma_{a1}=\gamma_{a3}=0$), and the owner admiration aspect(i.e., $\gamma_{a1}=\gamma_{a2}=0$). We present the learned attention weights of different aspects of our proposed HASC model in the fourth column. The last column gives some intuitive explanations of the experimental results. As shown in this figure, by learning the importance of different aspects with attentive modeling, HASC could better learn each user's preference from various social contextual aspects. Thus, it shows the the best performance for the users in the first three rows. In the fourth row, we present a case that all the models do not perform well expect than the simplified $C$ model from HASC that leverages the single creator admiration aspect into consideration. We carefully analyze this user's records  and guess a possible reason is that: the style and the content of the test image has rarely appeared in the user's  training data. As this test image differs from the distribution of the training images of this user, most models could not perform well. However, the $C$ model that leverages the owner admiration shows better results than the remaining models, as this user has liked several images uploaded by the owner. This example gives us an intuitive explanation that shows when our proposed model may not perform very well. Nevertheless, we must notice that this case is caused by the situation that the test pattern is not consistent with the patterns in the training data, which is uncommon. Therefore, we could empirically conclude that our proposed model shows the best results for most cases.

\vspace{-0.3cm}
\section{Conclusions}
In this paper, we have proposed a hierarchical attentive social contextual model of HASC for social contextual image recommendation. Specifically, in addition to user interest modeling, we have identified three social contextual aspects that influence a user's preference to an image from heterogeneous data: the upload history aspect, the social influence aspect, and the owner admiration aspect. We designed a hierarchical attention network that naturally mirrored the hierarchical relationship of users' interest given the three identified aspects. In the meantime, by feeding the
data embedding from rich heterogeneous data sources, the hierarchical attention networks could learn to attend differently to more or less important content. Extensive experiments on real-world datasets clearly demonstrated that our proposed HASC model consistently outperforms various state-of-the-art baselines for image recommendation.

%

\begin{small}
\bibliographystyle{abbrv}
\bibliography{kddimage2018}
\end{small}

\begin{IEEEbiography}[{\includegraphics[width=1in,height=1.25in,clip,keepaspectratio]{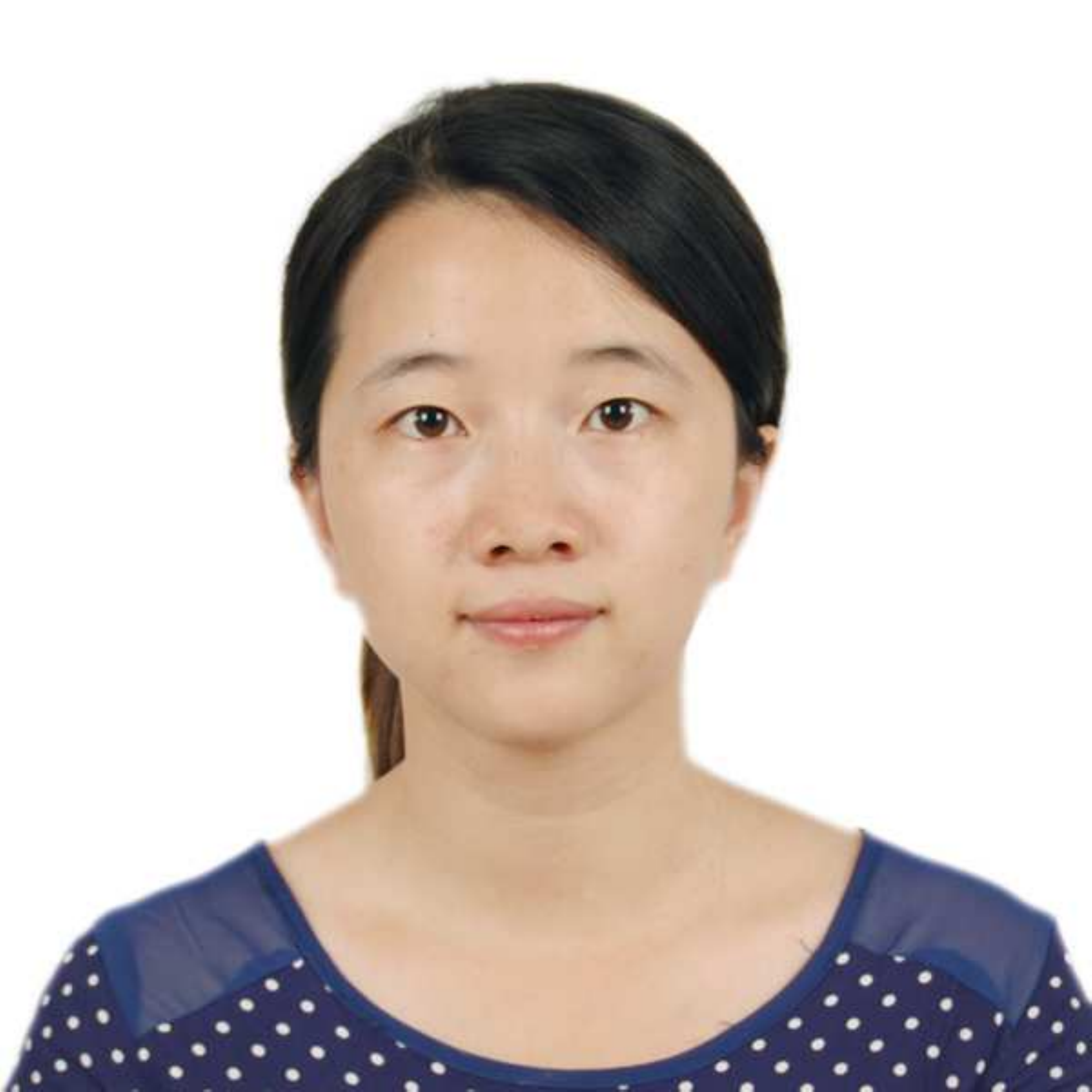}}]
{Le Wu} is currently an assistant professor at the Hefei University of Technology (HFUT), China. She received the Ph.D. degree from the University of Science and Technology of China (USTC). Her general area of research  interests is data mining, recommender systems and social network analysis. She has published more than 30 papers in referred journals and conferences. Dr. Le Wu is the recipient of the Best of SDM 2015 Award, and the Distinguished Dissertation Award from China Association for Artificial Intelligence (CAAI) 2017.
\end{IEEEbiography}

\begin{IEEEbiography}[{\includegraphics[width=1in,height=1.25in,clip,keepaspectratio]{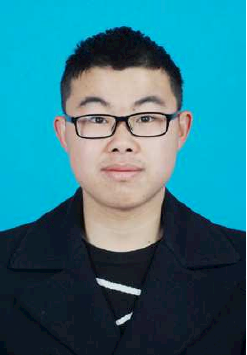}}]
{Lei Chen} is currently working towards the M.S. degree at Hefei University of Technology, China. He received the B.S. degree from Anhui University in 2016. His research interests include multimedia analysis and data mining.
\end{IEEEbiography}

\begin{IEEEbiography}[{\includegraphics[width=1.0in,height=1.25in,clip,keepaspectratio]{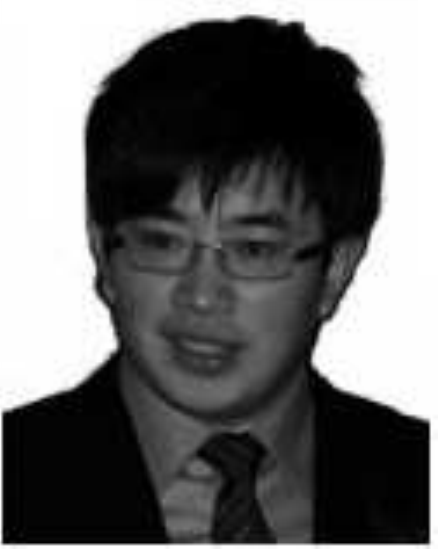}}]
{Richang Hong}(M'12) is currently a professor at HFUT. He  received the Ph.D. degree from USTC, in 2008. He has co-authored over 60 publications in the areas of his research interests, which include multimedia question answering, video content analysis, and pattern recognition. He is a member of the Association for Computing Machinery. He was a recipient of the best paper award in the ACM Multimedia 2010.
\end{IEEEbiography}

\begin{IEEEbiography}[{\includegraphics[width=1in,height=1.25in,clip,keepaspectratio]{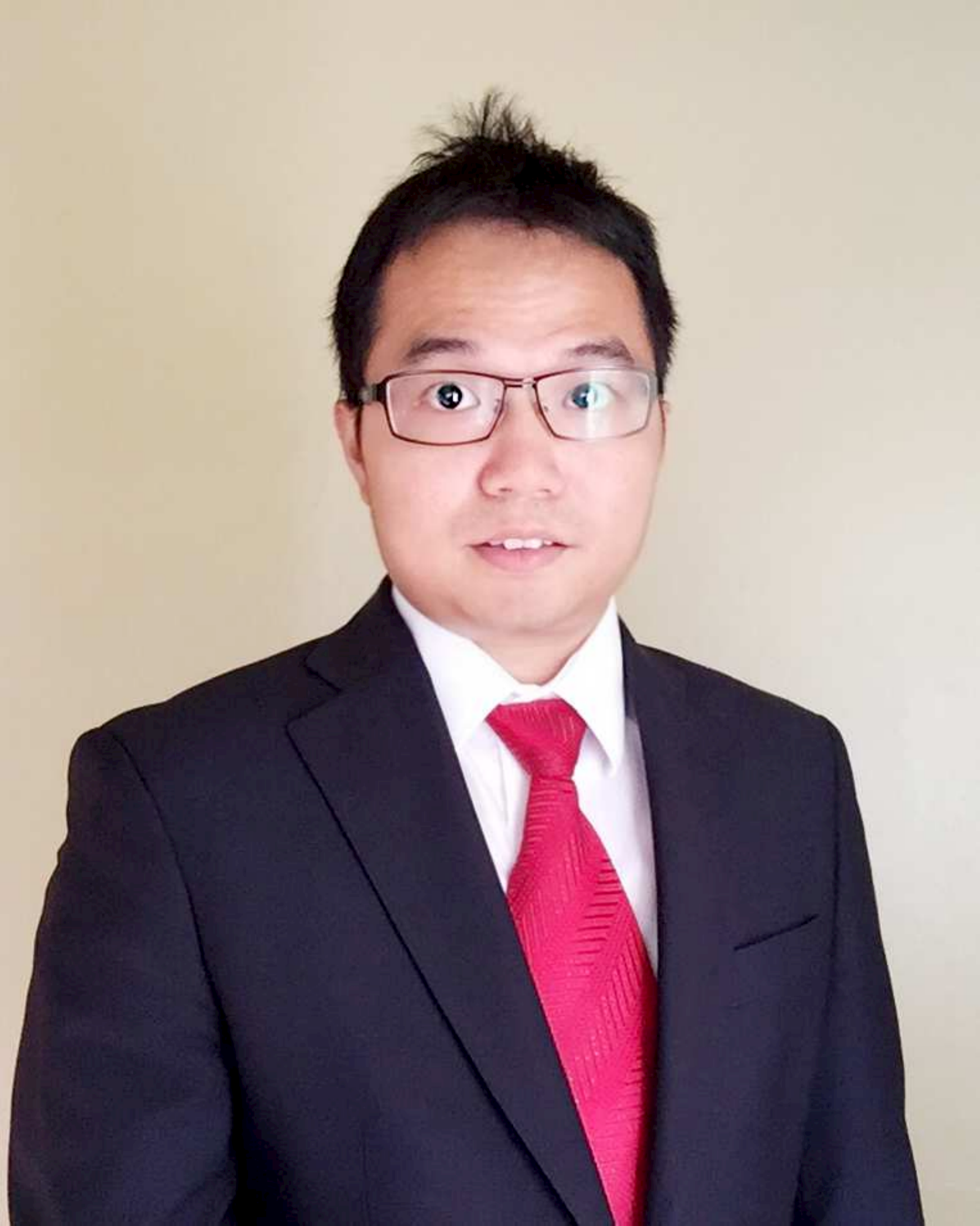}}]
{Yanjie Fu} received his Ph.D. degree from Rugters University in 2016, the B.E. degree from University of Science and Technology of China in 2008, and the M.E. degree from Chinese Academy of Sciences in 2011. He is currently an Assistant Professor at the Missouri University of Science and Technology. His general interests are data mining and big data analytics. He has published proficiently in referred journals and conference proceedings, such as IEEE TKDE, ACM TKDD, IEEE TMC and ACM SIGKDD.
\end{IEEEbiography}

\begin{IEEEbiography}[{\includegraphics[width=1in,height=1.25in,clip,keepaspectratio]{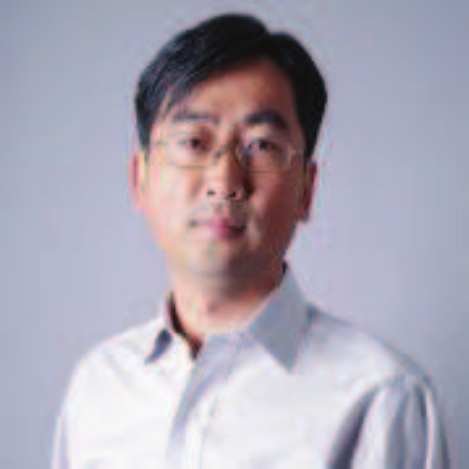}}]
{Xing Xie}(SM'09) is currently a senior researcher in Microsoft Research Asia, and a guest PhD advisor at USTC. His research interest include spatial data mining, location-based services, social networks, and ubiquitous computing. In recent years, he was involved in the program or organizing committees of over 70 conferences and works. Especially, he initiated the LBSN workshop series and served as program co-chair of ACM Ubicomp 2011. He is a senior member of ACM and the IEEE, and a distinguished member of China Computer Federation (CCF).
\end{IEEEbiography}

\begin{IEEEbiography}[{\includegraphics[width=1.0in,height=1.25in,clip,keepaspectratio]{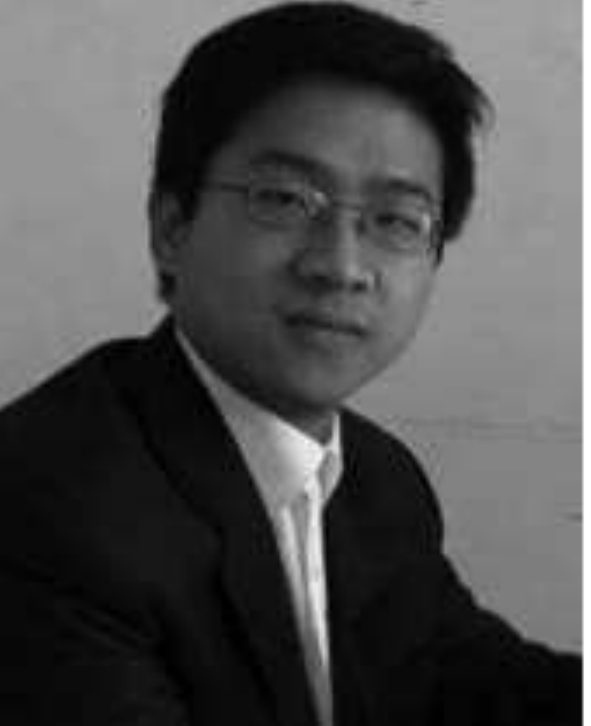}}]
{Meng Wang } is a professor at the Hefei University of Technology,
China. He received his B.E. degree and Ph.D. degree in the Special
Class for the Gifted Young and the Department of Electronic
Engineering and Information Science from the University of Science and
Technology of China (USTC), Hefei, China, in 2003 and 2008,
respectively. His current research interests include multimedia
content analysis, computer vision, and pattern recognition. He has
authored more than 200 book chapters, journal and conference papers in
these areas. He is the recipient of the ACM SIGMM Rising Star Award 2014.
He is an associate editor of IEEE Transactions on Knowledge and Data
Engineering (IEEE TKDE), IEEE Transactions on Circuits and Systems
for Video Technology (IEEE TCSVT), IEEE Transactions on Multimedia (IEEE TMM), and IEEE Transactions on Neural Networks and Learning Systems (IEEE TNNLS).
\end{IEEEbiography}

\end{document}